\documentclass[aps,preprint,superscriptaddress,groupedaddress, nofootinbib]{revtex4}  
\usepackage[dvipdfmx]{graphicx}  
\usepackage{dcolumn}   
\usepackage{bm}        
\usepackage{amssymb}   
\usepackage{amsmath}   
\usepackage{color}
\usepackage{ulem}

\begin{document}



\title{$I=2$ $\pi\pi$ scattering phase shift from the HAL QCD method with the LapH smearing}
\author{Daisuke~Kawai} \affiliation{Department of Physics, Kyoto University, Kyoto 606-8502, Japan}
\author{Sinya~Aoki} \affiliation{Center for Gravitational Physics, Yukawa Institute for Theoretical Physics, Kyoto University, Kyoto 606-8502, Japan}
\author{Takumi~Doi} \affiliation{Theoretical Research Division, Nishina Center, RIKEN, Saitama 351-0198, Japan}
\affiliation{iTHEMS Program and iTHES Research Group, RIKEN, Wako 351-0198, Japan}
\author{Yoichi~Ikeda} \affiliation{Research Center for Nuclear Physics (RCNP), Osaka University, Osaka 567-0047, Japan}
\author{Takashi~Inoue} \affiliation{Nihon University, College of Bioresource Sciences, Kanagawa 252-0880, Japan}
\author{Takumi~Iritani} \affiliation{Theoretical Research Division, Nishina Center, RIKEN, Saitama 351-0198, Japan}
\author{Noriyoshi~Ishii} \affiliation{Research Center for Nuclear Physics (RCNP), Osaka University, Osaka 567-0047, Japan}
\author{Takaya~Miyamoto} \affiliation{Center for Gravitational Physics, Yukawa Institute for Theoretical Physics, Kyoto University, Kyoto 606-8502, Japan}
\author{Hidekatsu~Nemura} \affiliation{Research Center for Nuclear Physics (RCNP), Osaka University, Osaka 567-0047, Japan}
\author{Kenji~Sasaki} \affiliation{Center for Gravitational Physics, Yukawa Institute for Theoretical Physics, Kyoto University, Kyoto 606-8502, Japan}
\collaboration{HAL~QCD~Collaboration}
\date{\today}

\begin{abstract}
Physical observables, such as the scattering phase shifts and the binding energies,
 calculated from the non-local HAL QCD potential do not depend on the sink operators used to  define the  potential.  This is called the 
 scheme independence of the HAL QCD method.
In practical applications,  the derivative expansion of the non-local potential  is employed, so that 
physical observables may receive some scheme dependence  at given order of the expansion.
In this paper, we compare   the $I=2$ $\pi\pi$ scattering phase shifts obtained in   
the point-sink scheme (the standard scheme in the HAL QCD method) and 
 the smeared-sink scheme (the LapH smearing newly introduced in the HAL QCD method).
Although potentials in different schemes have different forms as expected, 
we find that, for reasonably small smearing size, the resultant scattering phase shifts agree with each other   
 if  the next-to-leading order (NLO) term is taken into account.
 We also find that the HAL QCD potential in the point-sink scheme has negligible NLO term for wide range of energies, which
 implies  a good convergence of the derivative expansion in this case, while  the potential in the smeared-sink scheme has non-negligible NLO contribution.
Implication of this observation to the future studies of resonance channels (such as the $I=0$ and $1$ $\pi\pi$ scatterings)
with smeared all-to-all propagators is briefly discussed.
 All computations in this paper have been  performed at the lattice spacing $a\simeq 0.12$ fm ($1/a \simeq 1.6$ GeV) 
 on a $16^3\times 32$ lattice with the pion mass  $m_\pi\simeq 870$ MeV.
\end{abstract}

\preprint{KUNS-2706,~YITP-17-115,~RIKEN-QHP-337}
\pacs{11.15.Ha,12.38.Gc}
\maketitle

\section{Introduction}
Understanding the spectra of hadrons including resonant states
from the fundamental theory, quantum chromodynamics (QCD), is one of the major goals in
particle and nuclear physics.
Lattice QCD is a well-established approach as the first-principles calculation of QCD:
The calculation of the spectra of the ground state of the single hadron has been matured
and it is an important next challenge to determine interactions between hadrons,
which are essential to study the hadron-hadron scatterings and properties of
resonances such as $\rho$ and $\Delta$.
%
%
In lattice QCD, hadron-hadron interactions are mainly investigated by two methods.
The first one is the L\"uscher's finite volume method \cite{luscher} and its extensions \cite{moving_frame,coupled_channel1,coupled_channel2,coupled_channel3}.
In this approach, the energies of the states on finite volume lattices are extracted from temporal correlation functions
and are converted to the scattering phase shifts in the infinite volume through the L\"{u}scher's finite volume formula~\cite{luscher}.
This method has been applied to the meson-meson interactions not only  in non-resonant channels but also in resonant channels such as $\rho$, $a_0$ and $\sigma$~\cite{rho1, rho2, sigma1, sigma2, sigma3, a01, a02},
by using the advanced numerical techniques
including the variational method \cite{vari1,vari2} and all-to-all propagators.
The second method is the HAL QCD method \cite{HAL,HAL2, HAL3},
in which non-local but energy-independent potentials are extracted
from the space-time dependence of the Nambu-Bethe-Salpeter (NBS) wave functions.
Physical observables such as phase shifts and binding energies are then obtained
by solving the Schr\"{o}dinger equation using the HAL QCD potentials.
This method is also applied for wide range of
hadron systems \cite{spec1,spec2, spec3, spec4, spec5, spec6, spec7, spec8, spec9, spec10, spec11, three_body, lambda_c}
including the candidate of exotic tetraquark resonance, Z$_c$(3900)  \cite{zc3900}.
While HAL QCD method has been mainly used for channels in which
a so-called disconnected diagram is absent, 
it can be also applied for
systems with disconnected diagrams, which are typical for resonant channels,
by incorporating a method to calculate all-to-all propagators with an affordable computational cost.
In this work, we perform a lattice QCD calculation of $I=2$ $\pi\pi$ scatterings
from the HAL QCD method with the Laplacian Heaviside (LapH) method (or distillation) \cite{distill, stochastic}
to calculate all-to-all propagators,
as a first step toward the future studies of resonant states such as $\rho$ and $\sigma$.

With the use of the LapH method,
the operator is automatically smeared.
It is then of importance to study the dependence of results on the smearing of operators,
since the NBS wave function and the potentials are defined by the sink operators
in the HAL QCD method.
Theoretically, the physical observables extracted from non-local potentials
are independent of the definition of sink operators (such as smearing)
even though the form of potentials itself will vary depending on the operators \cite{spec2, nishijima1958,zimmermann1958-1987,haag1958}.
We call this fact as ``scheme independence of HAL QCD method''.
In the practical calculations, however, derivative expansion in terms of the non-locality of the potential is employed,
so that some sink operator dependence might appear in physical observables for a given order of the expansion.
In this study, we investigate the scheme independence in the HAL QCD method,
by comparing the phase shifts extracted from the potential defined
with point sink operator (``point-sink scheme''), which is standard operator in HAL QCD method,
with those defined with smeared sink operator (``smeared-sink scheme'').
In order to make a precise investigation, we consider the $I=2$ $\pi\pi$ scattering as a benchmark system.
%
The $I=2$ $\pi\pi$ scattering phase shifts are well-studied by experiments \cite{experi1,experi2, experi3}
and lattice calculations by using L\"{u}scher's finite volume method \cite{Dudek:1011,hadI2}.
Moreover, the calculation of NBS wave functions for this channel can be done in an affordable cost even
with point-sink scheme because the propagation of quarks in the same time slice is absent.
The consistency check between the HAL QCD method with the point sink operator
(without LapH smearing) and the L\"{u}scher's finite volume method in this channel has been performed  
in quenched QCD \cite{I2pre}, where
the good agreement of the phase shifts between the HAL QCD method and the L\"{u}scher's finite volume method has been demonstrated.
Thus, a purpose of this paper is to establish
the sink operator independence of the scattering phase shifts in this channel even when the potential is modified due to smeared sink operators.

This paper is organized as follows.
In Sec.~\ref{sec:scheme}, we introduce the HAL QCD potential method and explain the  LapH smearing scheme for the operator construction. 
In Sec.~\ref{sec:sec4}, the numerical setup used in this study is explained.
In Sec.~\ref{sec:sec3}, 
we calculate the potential in several different schemes, from which the scattering phase shifts are obtained.
This section is the main part of this paper, where we investigate the convergence of the derivative expansion in different sink operator schemes.
Sec.~\ref{sec:sec5} is devoted to the summary of this paper.
Some technical details are given in appendices. 

\section{The scheme in the HAL QCD method}
\label{sec:scheme}
\subsection{Original HAL QCD method}
The most important quantity in the HAL QCD method is the Nambu-Bethe-Salpeter (NBS) wave function, which is defined for the $I=2$ $\pi\pi$ system as
\begin{eqnarray}
\label{eq:NBS}
\psi^{W_k}_{n_s}(\bold{r}) & = \sum_{\bold{x}} &\left\langle 0 \vert \pi^{-}_{n_s}(\bold{x},0) \pi^{-}_{n_s}(\bold{x+r}, 0)\vert \pi^-\pi^-, W_k\right\rangle, 
\end{eqnarray}
where $\vert 0\rangle$ is the QCD vacuum, $\vert \pi^-\pi^-, W_k\rangle$ is  the $\pi\pi$ eigenstate 
in  the $(I, I_z)=(2, -2)$ channel,  
$W_k = 2 \sqrt{m_\pi^2 + k^2}$ is the central mass energy with the momentum $\bold{k}$ and its magnitude $k\equiv \vert \bold{k} \vert$, and
$\pi^-_{n_s}(x)$ is the negatively charged pion operator, defined by
\begin{eqnarray}
\pi^-_{n_s}(x) &=& \sum_a\bar u_{n_s}^a(x) \gamma_5 d_{n_s}^a(x),
\end{eqnarray}
where $a$ is the index for color, 
and the label $n_s$ represents the smearing level, which will be explained in the later subsection.
The crucial property of the NBS wave function (below inelastic threshold $W_{\rm th}=4m_\pi$) is that
the phase shift $\delta(k)$ is encoded in the asymptotic behaviors of $\psi^{W_k}_{n_s}(\bold{r})$~\cite{spec1,spec2,HAL2}.

The non-local but energy-independent potential is defined from the NBS wave function
as
\begin{eqnarray}
  \label{schrodinger}
\left[\frac{k^2}{m_\pi} -H_0 \right] \psi_{n_s}^{W_k}(\bold{r}) &=& \int d^3 r^\prime U_{n_s} (\bold{r}, \bold{r^\prime} ) \psi^{W_k}_{n_s}(\bold{r^\prime}), \quad{(W_k < W_{\rm th})}
\end{eqnarray}
where $H_0 \equiv -\nabla^2/m_\pi$, and $m_\pi$ is the pion mass.
The potential $U_{n_s} (\bold{r}, \bold{r^\prime} )$ is faithful to the phase shift below inelastic threshold,
while $U_{n_s} (\bold{r}, \bold{r^\prime} )$
itself is not a physical observable and
explicitly depends
on
the definition of the pion operator in NBS wave function,
such as the smearing level $n_s$.

\subsection{Time-dependent HAL QCD method}
In this subsection,  we explain  the time-dependent HAL QCD method \cite{HAL3}, which we employ
to extract the potential reliably.

We denote the 2-pt correlation functions as 
\begin{eqnarray}
\mathcal{C}^2_{n_a,n_b} (t,t_0) &=&\sum_{\bold{x,y}}\left\langle \pi_{n_a}^-(\bold{x},t) \pi_{n_b}^+(\bold{y},t_0) \right\rangle, 
\end{eqnarray}
where $n_a$ ($n_b$) is the smearing level of the sink (source) operator, and
$\pi^-=\bar u \gamma_5 d$, $\pi^+ =\bar d \gamma_5 u$.
The potentials are extracted from the 4-pt correlation function with various combinations of $n_a$ and $n_b$, which is given by
\begin{eqnarray}
\mathcal{C}^{4,A^+_1,1}_{n_a,n_b} (\bold{r},t;\left|\bold{P}\right|,t_0) &=& 
\sum_{\bold{x}}  \left\langle \pi_{n_a}^- (\bold{x},t) \pi_{n_a}^-(\bold{x+r},t) (\pi_{n_b}\pi_{n_b})^{A^+_1,1}_{2,2}(\left|\bold{P}\right|,t_0)  \right\rangle ,
\label{smeared4pt}
\end{eqnarray}
with
\begin{eqnarray}
(\pi_{n_s}\pi_{n_s})^{\Lambda,\mu}_{I,I_z}(\left|\bold{P}\right|,t) &=& \sum_{\bold{P} \atop \left|\bold{P}\right| : {\rm fix}} \sum_{I_1, I_2} \sum_{\bold{x},\bold{y}} 
 C^{\Lambda,\mu}(\bold{P}) D_{I_1, I_2}^{I,I_z} e^{-i \bold{P}\cdot\bold{x}} e^{i \bold{P}\cdot\bold{y}} \pi_{n_s}^{I_1}\left(\bold{x},t\right)  \pi_{n_s}^{I_2} \left(\bold{y},t\right) ,
\end{eqnarray}
where $C^{\Lambda,\mu}(\bold{P})$ is the Clebsch-Gordan coefficients  in the $\mu$-th component of the irreducible representation $\Lambda$ of the cubic group, and  
the relative momentum $\bold{P}$ is an element of $\{g \bold{P_0} \, | \, g \in O_h , \bold{P_0} = [n,0,0] $ in lattice unit with   $\,|\bold{P_0}| \leq 2\}$,
while 
$D_{I_1, I_2}^{I,I_z}$  is that from two pions 
with the $z$ component of the isospin $I_1$ and  $I_2$ to the two pion system with the total isospin $I$ and its $z$ component $I_z$.
Hereafter we exclusively take $\Lambda=A^+_1$ and $\mu=1$. 

The 4-pt correlation function in Eq.~(\ref{smeared4pt})
can be decomposed into a product of the NBS wave function $\psi_{n_a}^{W_k}(\bold{r})$ and the time-dependent part for each eigenstate as
\begin{eqnarray}
  \label{eq:4ptCorr}
\mathcal{C}^{4,A^+_1,1}_{n_a,n_b}(\bold{r},t;\left|\bold{P}\right|,t_0) = \sum_k A^{W_k}_{n_b} \psi_{n_a}^{W_k}(\bold{r})  e^{-W_k (t-t_0)} + \cdots,
\end{eqnarray}
where 
$A_{n_b}^{W_k} \equiv \left\langle \pi^-\pi^-, W_k \left\vert (\pi_{n_b}\pi_{n_b})^{A^+_1,1}_{2,2}(\left\vert \bold{P}\right|,0) \right\vert 0 \right\rangle$ is the overlap between the QCD eigenstate and the vacuum with the insertion of a two-pion operator, and
the ellipsis represents inelastic contributions, which become negligible at moderately large $t-t_0$ and
thus will be neglected in further discussions.
Here indices for the irreducible representation and the isospin are omitted in $A^{W_k}_{n_b}$ for simplicity.

The R-correlator,  defined by 
\begin{eqnarray}
R^{A^+_1,1}_{n_a,n_b}(\bold{r},t;\left|\bold{P}\right|,t_0) \equiv \mathcal{C}_{n_a,n_b}^{4,A^+_1,1}(\bold{r},t;\left|\bold{P}\right|,t_0)/ \left\{\mathcal{C}^2_{n_a,n_b} (t,t_0) \right\}^2,~~~
\end{eqnarray}
satisfies
\begin{eqnarray}
\label{eq:schrodinger_raw}
 \left(\frac{1}{4m_\pi} \frac{\partial^2}{\partial t^2} - \frac{\partial}{\partial t} - H_0 \right) R^{A^+_1,1}_{n_a,n_b}(\bold{r},t;\left|\bold{P}\right|,t_0) 
&=&    \int d^3r^\prime U_{n_a}\left(\bold{r},\bold{r'}\right) R^{A^+_1,1}_{n_a,n_b}(\bold{r^\prime},t;\left|\bold{P}\right|,t_0),
\end{eqnarray}
where the scheme dependence of the potential on the sink operator is explicit as $U_{n_a}$ \cite{HAL3}.
It is essential that all elastic states can be used to extract the non-local potential  $U_{n_a}\left(\bold{r},\bold{r'}\right)$,
and the ground state saturation is not required any more in this method. 
We note that, by construction,  $U_{n_a}\left(\bold{r},\bold{r'}\right)$
does not depend on quantities in the source operator such as the relative momentum $\vert\bold{P}\vert$ and the source smearing level $n_b$.

In practice, we approximate the non-local potential $U_{n_a}\left(\bold{r}, \bold{r}'\right)$ 
by the first few orders of the derivative expansion as
\begin{eqnarray}
U_{n_a}(\bold{r},\bold{r}') = \left\{V^{(0)}_{n_a}(r) + V^{(1)}_{n_a} (r) \nabla^2 + \mathcal{O}(\nabla^4) \right\} \delta^{(3)} \left(\bold{r}-\bold{r}'  \right) .~~~~
\end{eqnarray}
In this paper, we extract 
the potentials in the next-to-leading order (NLO) decomposition,
$ V^{(0)}_{n_a}(r)$ and $ V^{(1)}_{n_a} (r)$, by combining R-correlators with $\vert\bold{P}\vert = 0, 1$, neglecting $\mathcal{O}(\nabla^4)$ terms,
as explained in the next section.
We also define the (effective) leading order (LO) potential given by
\begin{eqnarray}
V^{\rm LO}_{n_a}(r;\vert\bold{P}\vert,n_b) &\equiv & \frac{ \left( \frac{1}{4m_\pi} \frac{\partial^2}{\partial t^2} -\frac{\partial}{\partial t} - H_0 \right) R^{A^+_1,1}_{n_a,n_b}(\bold{r},t;\vert\bold{P}\vert,t_0)}{R^{A^+_1,1}_{n_a,n_b}(\bold{r},t;\vert\bold{P}\vert, t_0)} \nonumber \\
&=& V^{(0)}_{n_a}(r) + V^{(1)}_{n_a}(r) \frac{ \nabla^2 R^{A^+_1,1}_{n_a,n_b}(\bold{r},t;\vert\bold{P}\vert,t_0)}{R^{A^+_1,1}_{n_a,n_b}(\bold{r},t;\vert\bold{P}\vert,t_0)}, 
\label{eq:effLO}
\end{eqnarray}
where
we make relative momentum ($\vert\bold{P}\vert$) and source operator ($n_b$)  dependence, 
introduced by the second term, explicit as $V^{LO}_{n_a}(r;\vert\bold{P}\vert,n_b)$.
Thus,
if $V^{\rm LO}_{n_a}(r,\vert\bold{P}\vert, n_b)$ does not strongly depend  on $\vert\bold{P}\vert$ or $n_b$, we can conclude that
\begin{eqnarray}
 V^{(1)}_{n_a}(r) \frac{ \nabla^2 R^{A^+_1,1}_{n_a,n_b}(\bold{r},t;\vert\bold{P}\vert,t_0)}{R^{A^+_1,1}_{n_a,n_b}(\bold{r},t;\vert\bold{P}\vert,t_0)}
\end{eqnarray}
is small, so that 
$V^{\rm LO}_{n_a}(r;\vert\bold{P}\vert, n_b) ( \simeq V^{(0)}_{n_a}(r) )$ can be used to calculate the scattering phase shifts reliably
around the energies probed by the R-correlators.

\subsection{Phase shift}
Once the local potentials $V^{(i)}_{n_a}(r)$  are obtained, the scattering phase shifts can be calculated by solving the Schr\"{o}dinger equation, Eq.~(\ref{schrodinger}).
As noted before,
the phase shifts should be independent of
the sink operator scheme ($n_a$) 
as long as a sufficient number of local potentials $V^{(i)}_{n_a}$ $(i=0,1,\cdots)$ are employed
to represent the non-local potential, though each term $V^{(i)}_{n_a}$ might have sizable dependence on the sink operator scheme \cite{spec2}.

In this paper, we check this scheme independence of the $I=2$ $\pi\pi$ scattering phase shifts at the NLO level.
As it has already been shown~\cite{I2pre} and is confirmed in this paper (Sec.~\ref{subsec:point}), the contribution from NLO or higher order terms to
the potential  in the point-sink scheme is negligibly small in this channel, so that
the effective LO potential gives the correct phase shifts below a certain energy. 
Therefore, we can use the scattering phase shifts from the potential in the point-sink scheme as the benchmark of our analysis.
We compare the scattering phase shifts calculated from the potentials in the LapH smearing scheme with the benchmark, in order to see  
how good the NLO analysis in the derivative expansion is in this scheme.

\subsection{Some remarks and comments on Ref.~\cite{Yamazaki:2017gjl}}
As already stressed many times before \cite{spec1,spec2,HAL2},
the quantitative comparison between different schemes can be done
only through physical observables but not through potentials.
The comparison between potentials
is analogous to the comparison of  the running couplings
among different schemes such as $\alpha_{\overline{\rm MS}}(q)$ ($\overline{\rm MS}$ scheme), 
$\alpha_{\rm V}(q)$ (potential scheme) or  $\alpha_{\rm SF}(q)$ (Shr\"odinger functional scheme)
in QCD.
We must compare physical quantities such as the scattering phase shifts in the case of the potential or the scattering amplitudes in the case of the running coupling.
Although physical observables are scheme independent in principle, approximations introduced to calculate them bring the scheme dependence into observables.
An example for the approximations is
the truncation of the derivative expansion for the potential or that of the perturbative expansion for  the running coupling.
In such cases, one scheme is better than others for the fast convergence of the approximation. 
In the case of the potential, 
the point-sink scheme is shown to be a good scheme for the fast convergence of the derivative expansion \cite{spec6,I2pre}, so that even the local potential at the LO gives reasonable results at low energies. Analogous scheme dependence for the convergence of the perturbative expansion 
exist for the running coupling. 

Here we make a few comments on a recent paper \cite{Yamazaki:2017gjl} whose discussions are trivially invalidated as shown below.

The first point discussed in~\cite{Yamazaki:2017gjl} is the
relation between the (local) energy-dependent potential and phase shifts,
where it was claimed that the energy-dependent potential defined at a given energy
can give the correct phase shift only at that energy, but gives incorrect phase shifts at different energies.
We note that such a claim is nothing to do with the HAL QCD method,
since the theoretical formulation of the HAL QCD method is based
on not the energy-dependent potential,
but the non-local energy-independent potential $U(\bold{r}, \bold{r}^\prime)$,
where it was proven that the latter is faithful
to the phase shifts at all energies below the inelastic threshold~\cite{spec1,spec2,HAL2}.
As discussed above as well as in our previous papers~\cite{spec1,spec2,HAL2},
the derivative expansion for the non-local potential, so far employed
in our applications, gives some truncation errors at given orders of the expansion,
which however should not be misunderstood as the theoretical limitation of the HAL QCD method.

The second point discussed in Ref.~\cite{Yamazaki:2017gjl} is the derivative expansion of the non-local potential $U(\bold{r}, \bold{r}^\prime)$.
We first note that
the information of $\psi^{W_k}_{n_s}(\bold{r})$ (in particular the phase shifts $\delta(k)$) below the inelastic threshold
are encoded in $U(\bold{r}, \bold{r}^\prime)$,
where the degrees of freedom of $\bold{r}$ and $\bold{k}$ in the former
are implicitly converted to those of $\bold{r}$ and $\bold{r'}$ in the latter through Eq.~(\ref{schrodinger}).
The derivative expansion of the non-local potential is 
given by
\begin{eqnarray}
U(\bold{r}, \bold{r}^\prime) &=& \sum_{\bold{n}} V_{\bold{n}}(\bold{r})\nabla_\bold{r}^\bold{n}
\delta^{(3)}(\bold{r} - \bold{r}^\prime),
\quad  \nabla_\bold{r}^\bold{n} \equiv \nabla_{r_x}^{n_x} \nabla_{r_y}^{n_y} \nabla_{r_z}^{n_z},
\end{eqnarray}
where no symmetry is assumed. Ref.~\cite{Yamazaki:2017gjl} claimed that $V_{\bold{n}}(\bold{r})$
cannot be $\bold{k}$ independent  since $U(\bold{r}, \bold{r}^\prime)$ needs the same number of the degrees of freedom, $\bold{r}$ and $\bold{r}^\prime$, to keep the $\bold{k}$ independence of $U(\bold{r}, \bold{r}^\prime)$. Clearly this statement is incorrect since $\bold{n}$ has enough degrees of freedom to describe $\bold{r}^\prime$ dependence.
For instance, using the Taylor expansion as
\begin{eqnarray}
\int d^3 r^\prime \, U(\bold{r}, \bold{r}^\prime) \psi(\bold{r}^\prime; \bold{k}) &=& \sum_{\bold{n}}  
\int d^3 r^\prime U(\bold{r}, \bold{r}^\prime)
\frac{ (\bold{r}^\prime - \bold{r})^\bold{n}}{ \bold{n} !}
 \nabla_\bold{r}^\bold{n}\psi(\bold{r};\bold{k}), \\
\frac{ (\bold{r}^\prime - \bold{r})^\bold{n}}{ \bold{n} !} &\equiv&  \frac{(r^\prime-r)_x^{n_x}}{n_x !} \frac{(r^\prime-r)_y^{n_y}}{n_y !} \frac{(r^\prime-r)_z^{n_z}}{n_z !} , \nonumber
 \end{eqnarray}
we can express $V_{\bold{n}}(r)$ in terms of $U(\bold{r}, \bold{r}^\prime)$ as
\begin{eqnarray}
V_\bold{n}(\bold{r}) &=&  \int d^3 r^\prime U(\bold{r}, \bold{r}^\prime)\frac{ (\bold{r}^\prime - \bold{r})^\bold{n}}{ \bold{n} !}, 
\end{eqnarray}
which are manifestly $\bold{k}$ independent.

The third point discussed in Ref.~\cite{Yamazaki:2017gjl}
is a relation between the scattering phase shift and the smearing of operator.
Ref.~\cite{Yamazaki:2017gjl} considered the following relation (Eq. (11) in Ref.~\cite{Yamazaki:2017gjl} and called the ``fundamental relation" in that paper )
\begin{eqnarray}
-\int d^3r h(r;k) e^{-i \bold {k}\cdot \bold{r}} &=& \frac{4\pi}{k}e^{i\delta_0(k)} \sin\delta_0(k), 
\quad
h(r;k) \equiv (\nabla^2 + k^2) \psi_0(r;k),
\end{eqnarray}
where $\delta_0(k)$ is the S-wave scattering phase shift and $\psi_0$ is the S-wave NBS wave function.
One can further consider the smeared NBS wave function defined by
\begin{eqnarray}
  \label{eq:smeared-NBS}
  \tilde \psi_0(r;k) &=& \int d^3r^\prime s(\vert \bold{r}-  \bold{r}^\prime \vert ) \psi_0(r^\prime;k) ,
  \quad
 \tilde h(r;k) \equiv  (\nabla^2 + k^2) \tilde \psi_0(r;k),
\end{eqnarray}
with $s(r)$ is a smearing function.
In Ref.~\cite{Yamazaki:2017gjl},
it was claimed that the ``fundamental relation" does not hold by this smearing
as (Eq.~(26) in Ref.~\cite{Yamazaki:2017gjl})
\begin{eqnarray}
-\int d^3r  \tilde h(r;k)
 e^{-i \bold {k}\cdot \bold{r}} &=& -\int d^3r
\int d^3r^\prime  s(\vert \bold{r}-  \bold{r}^\prime \vert ) h(r^\prime;k) e^{-i \bold{k} \cdot \bold{r}}
\not=  \frac{4\pi}{k}e^{i\delta_0(k)} \sin\delta_0(k),
\end{eqnarray}
and the correct phase shift $\delta_0(k)$ is not obtained from the smeared NBS wave function, Eq.~(\ref{eq:smeared-NBS}).
This claim is also incorrect.
Using $\bold {k}\cdot \bold{r} = \bold {k} \cdot (\bold{r} - \bold{r}^\prime) +  \bold {k} \cdot \bold{r}^\prime$, we have
\begin{eqnarray}
-\int d^3r  \tilde h(r;k) e^{-i \bold{k} \cdot \bold{r}}  &=& - C(k) \int d^3r' h(r';k) e^{-i \bold{k}\cdot \bold{r}'}
= C(k) \frac{4\pi}{k}e^{i\delta_0(k)} \sin\delta_0(k),\\
C(k) &\equiv& \int d^3 r\, s(\vert \bold{r}\vert) e^{-i \bold {k}\cdot \bold{r}}, 
\end{eqnarray}
so that the ``fundamental relation"  is satisfied if we use $\tilde h(r;k)/C(k)$ instead of $\tilde h(r;k)$. This normalization is indeed necessary and correct  since
$\psi_0(r;k) =  j_0( kr) + \mbox{scattering wave}$ implies $
 \tilde \psi_0(r;k) = C(k) ( j_0( kr) + \mbox{scattering wave})$.

\subsection{LapH smearing}
\label{sec:sec2}
The smeared pion operator at time $t$ is constructed as
\begin{eqnarray}
\label{eq:s_pion}
\pi^{fg}_{n_s}(\bold{x},t) & = &\sum_a \bar{q}_{n_s}^{a,f}(\bold{x},t) \gamma_5 q_{n_s}^{a,g}(\bold{x},t)
\end{eqnarray}
from the smeared quark operator given by 
\begin{eqnarray}
q_{n_s}^{a,f}(\bold{x},t) & = & \sum_{b,\bold{y}}\mathcal{S}_{n_s}^{ab}(\bold{x},\bold{y};t) q^{b,f}(\bold{y},t),
\end{eqnarray}
where $q^{a,f}(\bold{x},t)$ is a local quark field with a color index $a$ and a flavor index $f$ ($f=1,2$ for $u,d$ quarks), $\mathcal{S}$ is a smearing operator at time $t$ with smearing level $n_s$ \cite{distill, stochastic}.
Note that the spinor indices of quarks are implicit and are summed over in the pion operator.
Hereafter a summation over repeated indices is assumed, unless otherwise stated.

In this paper, we employ the gauge covariant smearing operator $\mathcal{S}$, which is constructed from a gauge covariant lattice Laplacian at $t$ defined by
\begin{eqnarray}
\label{eq:laplacian}
  \widetilde{\Delta}^{ab}(\bold{x},\bold{y},t) &=& 
  \sum_{i=1}^{3}  \left(\widetilde{U}_{\hat{i}}^{ab}(\bold{x},t) \delta_{\bold{x},\bold{y}+\hat{i}} - 2 \delta_{\bold{x},\bold{y}} + \widetilde{U}_{\hat{i}}^{\dagger ab}(\bold{y},t)\delta_{\bold{x},\bold{y}-\hat{i}} \right), \nonumber
\end{eqnarray}
where $\widetilde{U}^{ab}_i(\bold{x},t)$ represents a stout-smeared link variable \cite{stout}.
This operator can be diagonalized as 
\begin{eqnarray}
\widetilde{\Delta}^{ab}(\bold{x},\bold{y},t) = \sum_{n=1}^{n_{\rm max}} V^{a}_{n}(\bold{x},t) \lambda_{n} V^{b\dagger}_n(\bold{y},t),
\end{eqnarray}
where $\lambda_n$ is the $n$-th eigenvalue with $\vert \lambda_n\vert  \le \vert \lambda_m\vert$ for $ n < m$,  $V_n^a(\bold{x},t)$ is the corresponding eigenvector,
and $n_{\rm max} = N_{\rm color} N_x N_y N_z$.
Using eigenvectors, $\mathcal{S}$ with the smearing level $n_s$ is given by
\begin{eqnarray}
\mathcal{S}_{n_s}^{ab}(\bold{x},\bold{y};t) = \sum_{n=1}^{n_s} V_{n}^{a}(\bold{x},t) V_n^{b\dagger}(\bold{y},t) ,
\end{eqnarray}
which is the projection operator to the space spanned by $n_s$ eigenvectors.
We call this smearing the LapH smearing with the level $n_s$.
Since the eigenmodes corresponding to larger values of $\vert\lambda_n\vert$ are absent in this subspace, the smearing operator $\mathcal{S}$ removes  high momentum components of quark fields in a gauge covariant manner.
Roughly speaking, $n_s=8, 16, 32$ and 64 in our setup correspond to the momentum cutoff of 680 MeV, 770 MeV, 900 MeV and 1100 MeV, respectively.
Note that  the point quark with no smearing is given by  $n_s=n_{\rm max}$.
We also study the wall quark operator (only at the source), for which we use the short-hand notation ``$n_s=0$''
  even though it does not belong to the LapH smearing.
 
\section{Numerical setup}
\label{sec:sec4}

Since  the main purpose of this paper is to investigate  the scheme independence of the scattering phase shifts,
we perform a lattice QCD calculation at a single heavy pion mass on a small lattice.
We  employ  the 2 + 1 flavor gauge configuration on a $16^3 \times 32$ lattice,
generated by JLQCD and CP-PACS collaborations
with a renormalization-group improved gauge action at $\beta = 1.83$ and non-perturbatively improved clover action with $c_{SW} = 1.7610$ at the hopping parameters $k_{ud} = 0.1376$ and $k_s = 0.1371$ \cite{conf1,conf2}.
These parameters correspond to the lattice spacing  $a \simeq 0.1214$ fm ($a^{-1} \simeq 1.625$ GeV), so that the physical lattice size is  $L^3\times T \simeq \left(1.94 \mbox{ fm} \right)^3 \times 3.88 \mbox{ fm}$, and 
the pion mass $m_\pi \simeq 870$ MeV.
The stout-smearing parameters for the link variables in the gauge covariant Laplacian operator are chosen as staple weight $\rho = 0.1$ and iteration number $n_\rho = 10$.
For the calculation of the NBS wave function,
the periodic boundary condition is employed in all directions,
except for the point sink and  smeared/wall source combinations,
where the Dirichlet boundary condition is employed in the temporal direction.
In the case of  the Dirichlet boundary condition, the source is located at the midpoint of the temporal direction.
In the case of the wall source, the Coulomb gauge fixing is employed.
The number of configurations used in this paper is 60 for all the sink-source combinations except for the point sink and wall source combination, where 700 configurations are used,
and we calculate with 32 different source time slices per configuration. 
We take $t-t_0 = 11$ for all cases, except $t-t_0=12$ for $n_a=n_b= 16$ case and the point-sink cases,
where the single pion 2-pt function is saturated by the ground state.

\section{Results}
\label{sec:sec3}

\subsection{Potentials and scattering phase shifts in the point-sink scheme}
\label{subsec:point}

\begin{figure}[htbp]
\includegraphics[clip, width=\textwidth]{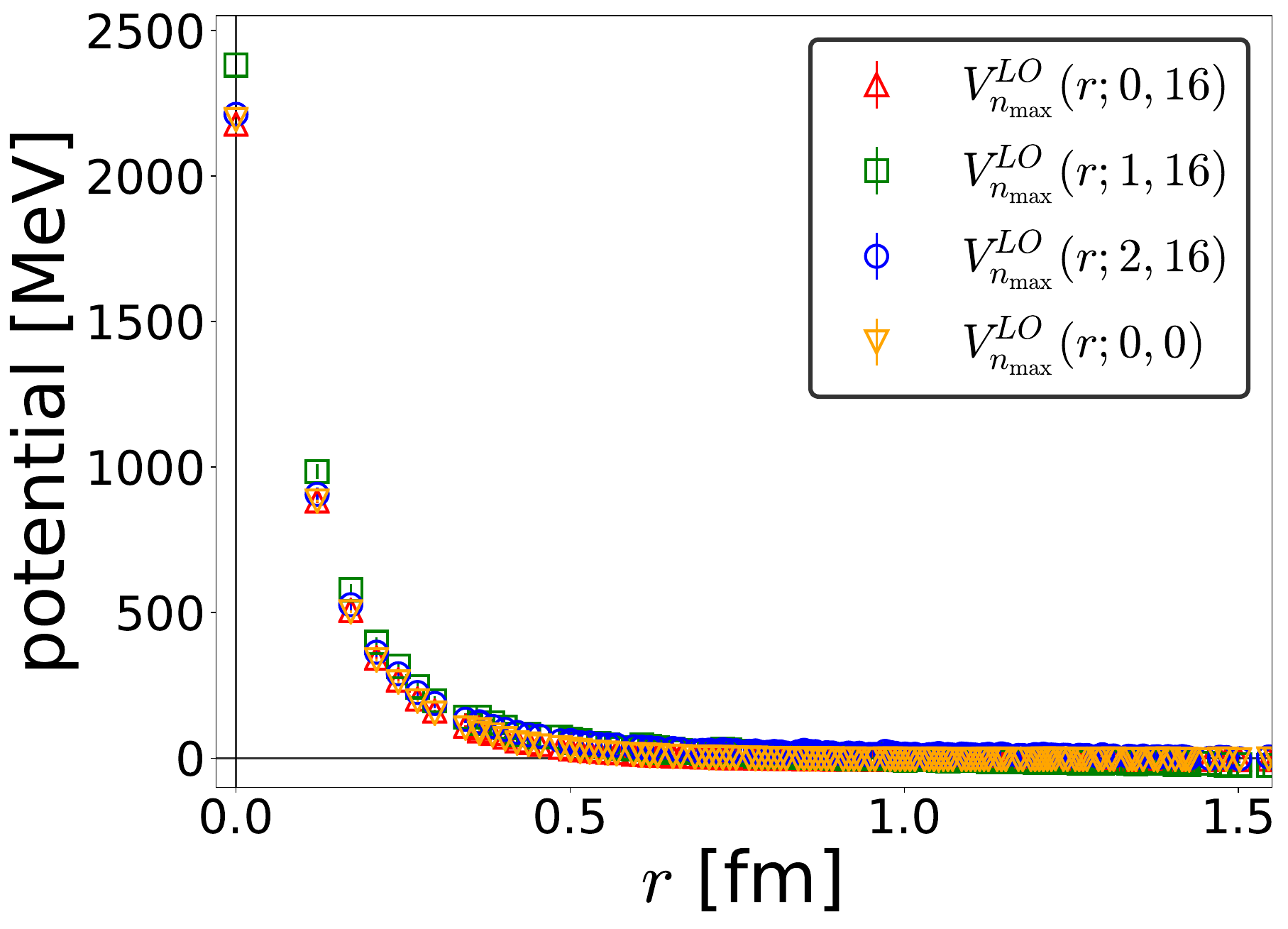}
\caption{The effective LO potentials in the point-sink scheme $V^{\rm LO}_{n_{\rm max}}(r;\vert \bold{P} \vert, n_b)$ with $n_b = 16$ 
  at $\vert \bold{P}\vert =$ 0 (red up triangles), 1(green squares), 2(blue circles), together with $V^{\rm LO}_{n_{\rm max}}(r; 0, n_b)$ with $n_b = 0$ (orange down triangles).
}
\label{fig:tot_pot_psink}
\end{figure}

We first study the point-sink scheme.
In Fig.~\ref{fig:tot_pot_psink}, we show the effective LO potentials in the point-sink scheme, $V_{n_{\rm max}}^{\rm LO}(r; \vert \bold{P} \vert, 16)$
with $\vert \bold{P} \vert$ = 0 (red up triangles), 1 (green squares), 2 (blue circles),
together with the one calculated from the wall source, $V_{n_{\rm max}}^{\rm LO}(r; \vert \bold{P} \vert =0 , 0)$ (orange down triangles).
The temporal separation, $t-t_0 = 12$, is large enough for the potential to be stable against the change of  $t-t_0$.
We first notice that the source operator dependence, the difference between the smeared with $n_b=16$ and the wall source is negligible, by comparing $V_{n_{\rm max}}^{\rm LO}(r; \vert \bold{P} \vert = 0, 16)$  (red up triangles) and $V_{n_{\rm max}}^{\rm LO}(r; \vert \bold{P} \vert =0 , 0)$ (orange down triangles).
In addition, it is observed that
$V_{n_{\rm max}}^{\rm LO}(r; \vert \bold{P} \vert,16)$ is almost independent of the source momentum $\vert \bold{P} \vert \leq 2$.

\begin{figure}[htbp]
\begin{center}
\includegraphics[clip, width=0.85\textwidth]{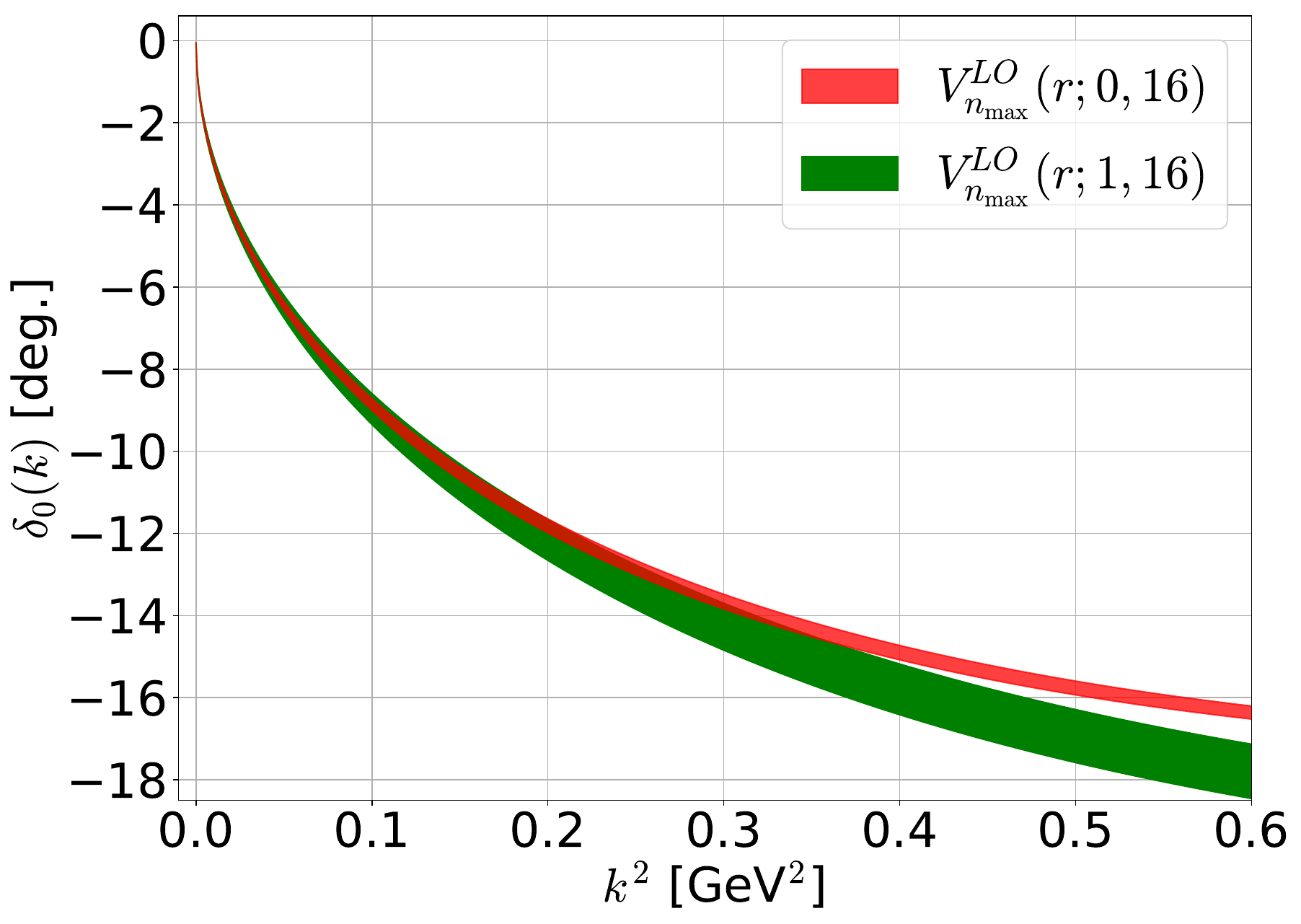}
\end{center}
\begin{center}
\includegraphics[clip, width=0.85\textwidth]{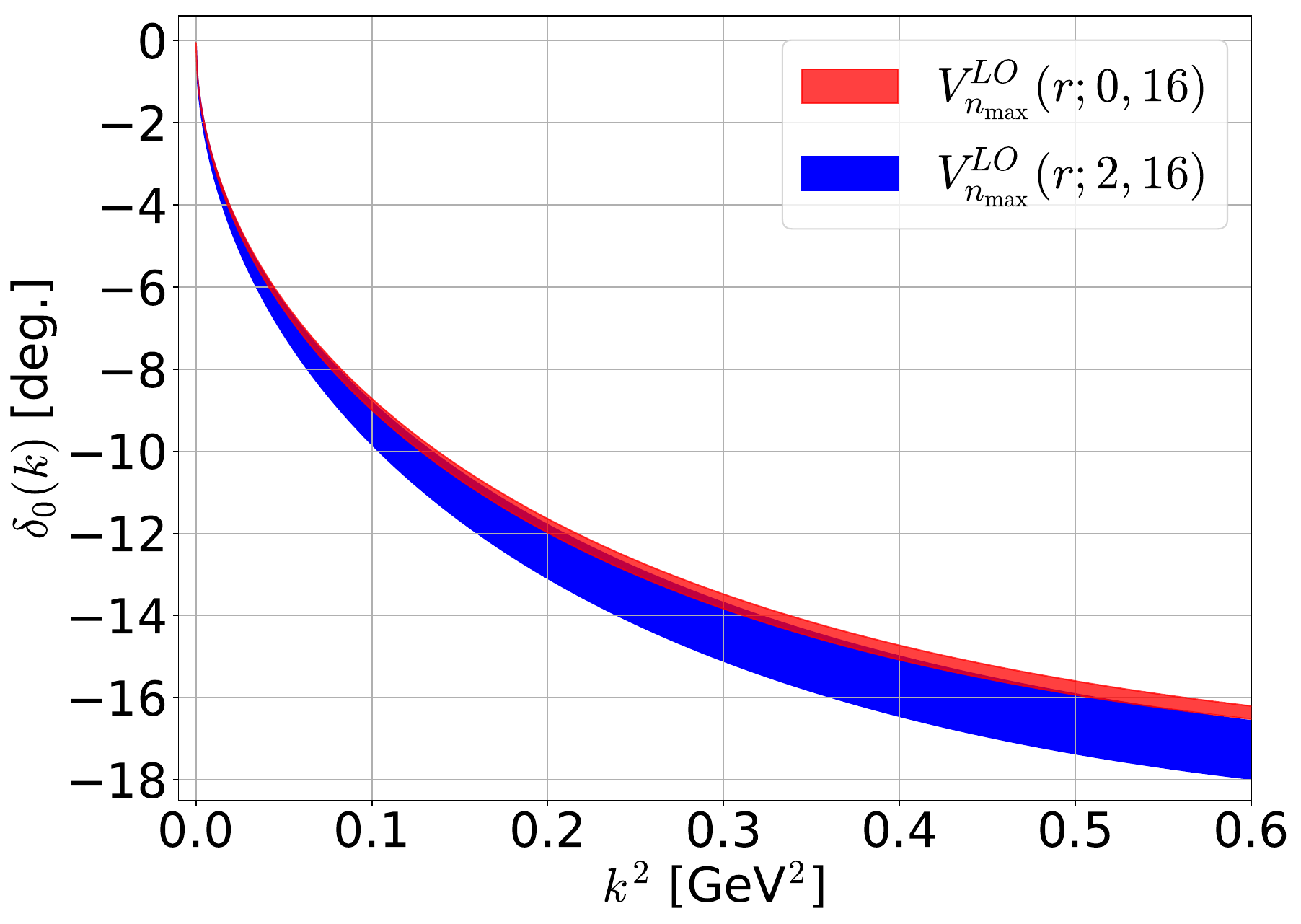}
\end{center}
\caption{
\label{fig:tot_phase_psink}
The phase shifts of the S-wave $I=2$ $\pi\pi$ scattering in the point-sink scheme
$V^{\rm LO}_{n_{\rm max}}(r;\vert \bold{P} \vert, 16)$.
(Upper) 
The results from
$\vert \bold{P} \vert = 0$ (red) and
$\vert \bold{P} \vert = 1$ (green) are compared.
(Lower)
The results from
$\vert \bold{P} \vert = 0$ (red) and
$\vert \bold{P} \vert = 2$ (blue) are compared.
}
\end{figure}

We obtain the scattering phase shifts
by solving the Schr\"{o}dinger equation with the potentials
which are fit by the following functional form,
\begin{equation}
  V^{\rm LO}(r) = A_1 e^{-\frac{r^2}{\sigma_1^2}} + A_2 e^{-\frac{r^2}{\sigma_2^2}} .
  \label{eq:pot_effLO}
\end{equation}
The fit of potential
works well for $\vert \bold{P} \vert = 0$ and $2$,
while there exist some residual errors for $\vert \bold{P} \vert = 1$,
since $V_{n_{\rm max}}^{\rm LO}(r; \vert \bold{P} \vert = 1, 16)$ at large $r$ slightly deviates from zero.
This introduces systematic uncertainties in results from $\vert \bold{P} \vert = 1$,
as discussed later.

Shown in Fig.~\ref{fig:tot_phase_psink}
are the obtained phase shifts in terms of $k^2$, where
we compare the results from $V_{n_{\rm max}}^{\rm LO}(r; \vert \bold{P} \vert, 16)$
between $\vert \bold{P} \vert = 0$ and $1$ (upper panel)
and
between $\vert \bold{P} \vert = 0$ and $2$ (lower panel),
where the bands correspond to the statistical errors only.
The phase shifts from 
$V_{n_{\rm max}}^{\rm LO}(r; \vert \bold{P} \vert = 0, 0)$ are almost identical
to those from $V_{n_{\rm max}}^{\rm LO}(r; \vert \bold{P} \vert = 0, 16)$,
and are not shown in the figure.
The most important observation in Fig.~\ref{fig:tot_phase_psink} is that,
at low energies,
the phase shifts are independent of the source within statistical errors
and the effective LO potential in the point-sink scheme describes the $I=2$ $\pi\pi$
scattering rather precisely.
We thus confirm that the conclusion of the previous quenched study for the $I=2$ $\pi\pi$ 
scattering \cite{I2pre} that the potential in the point-sink scheme, the standard for the HAL QCD potential,
is a good scheme also in the full QCD, so that the effective LO potential gives
a reliable approximation for the non-local potential up to a certain energy.
In order to estimate the energy above which NLO corrections would affect the results,
we look at Fig.~\ref{fig:tot_phase_psink} in more details.
In the upper panel,
we find that the results from $\vert \bold{P} \vert = 0$ (red)
and $\vert \bold{P} \vert = 1$ (green) are consistent with each other
at least up to $k^2 \lesssim 0.4$ GeV$^2$.
As discussed above, the results from $\vert \bold{P} \vert = 1$ would suffer from
additional systematic errors,
and thus the results from $\vert \bold{P} \vert = 0$ and $1$ may be consistent
even at higher energies.
In the lower panel,
we find that the results from $\vert \bold{P} \vert = 0$ (red)
and $\vert \bold{P} \vert = 2$ (blue) are consistent with each other
at least up to $k^2 \lesssim 0.6$ GeV$^2$.
As a conservative estimate,
we conclude that effective LO analysis in the point-sink scheme is reliable at least up to $k^2 \lesssim 0.4$ GeV$^2$.
Among the results in the point-sink scheme,
those from $\vert \bold{P} \vert = 0$ give the smallest statistical fluctuations,
and we therefore use the scattering phase shifts from  $V_{m_{\rm max}}^{\rm LO}(r; \vert \bold{P} \vert=0,16)$
as the benchmark in our investigation hereafter.

\subsection{Potentials in the smeared-sink scheme}
\label{subsec3}
\begin{figure}[htbp]
\includegraphics[clip, width=\textwidth]{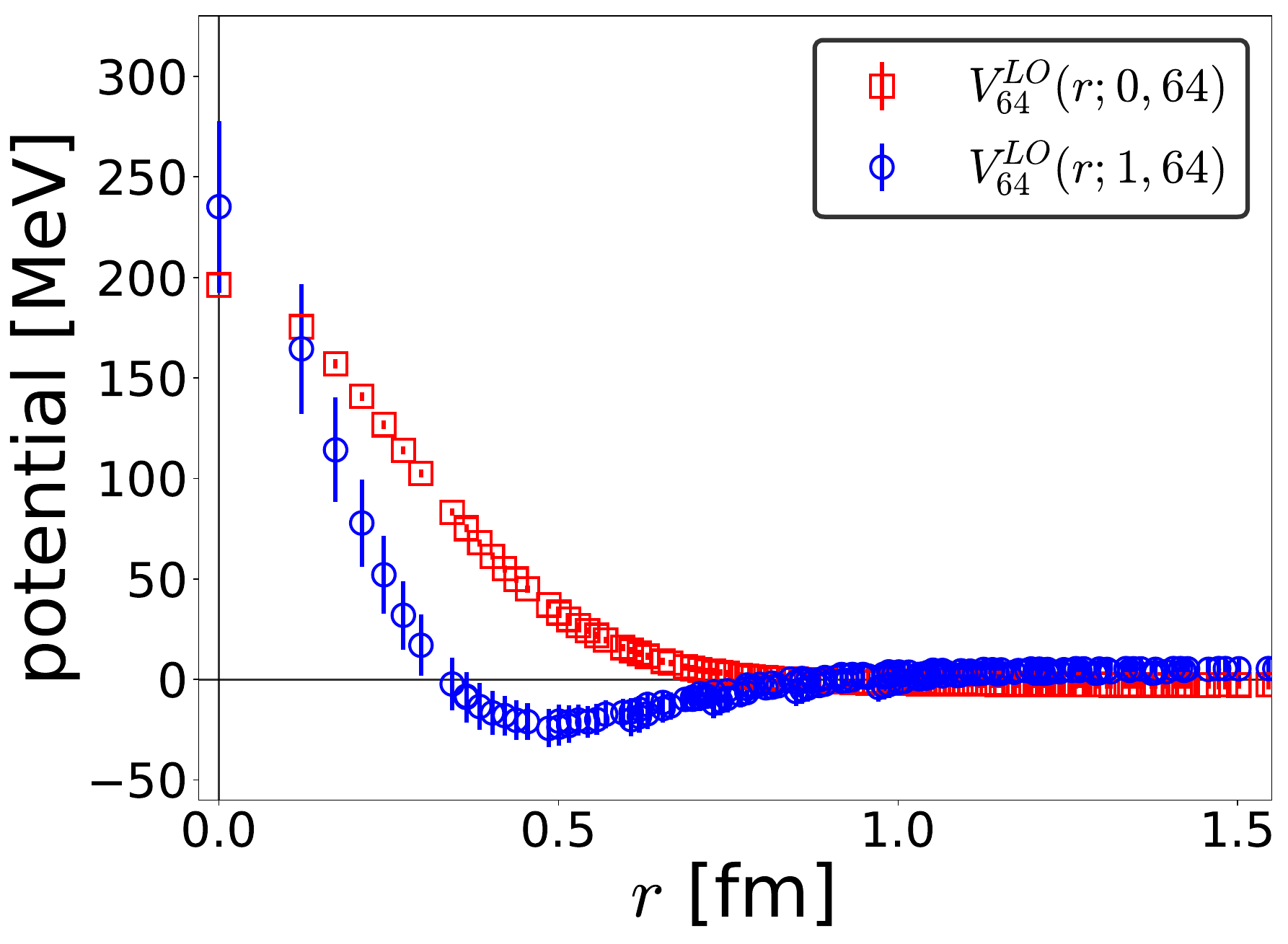}
\caption{The effective LO potential from the smeared-sink scheme $V^{\rm LO}_{64}(r;\vert\bold{P}\vert,64)$ 
at $\vert \bold{P}\vert = 0,1$.}
\label{fig:tot_pot_ssink}
\end{figure}

\begin{figure}[htbp]
\includegraphics[clip, width=\textwidth]{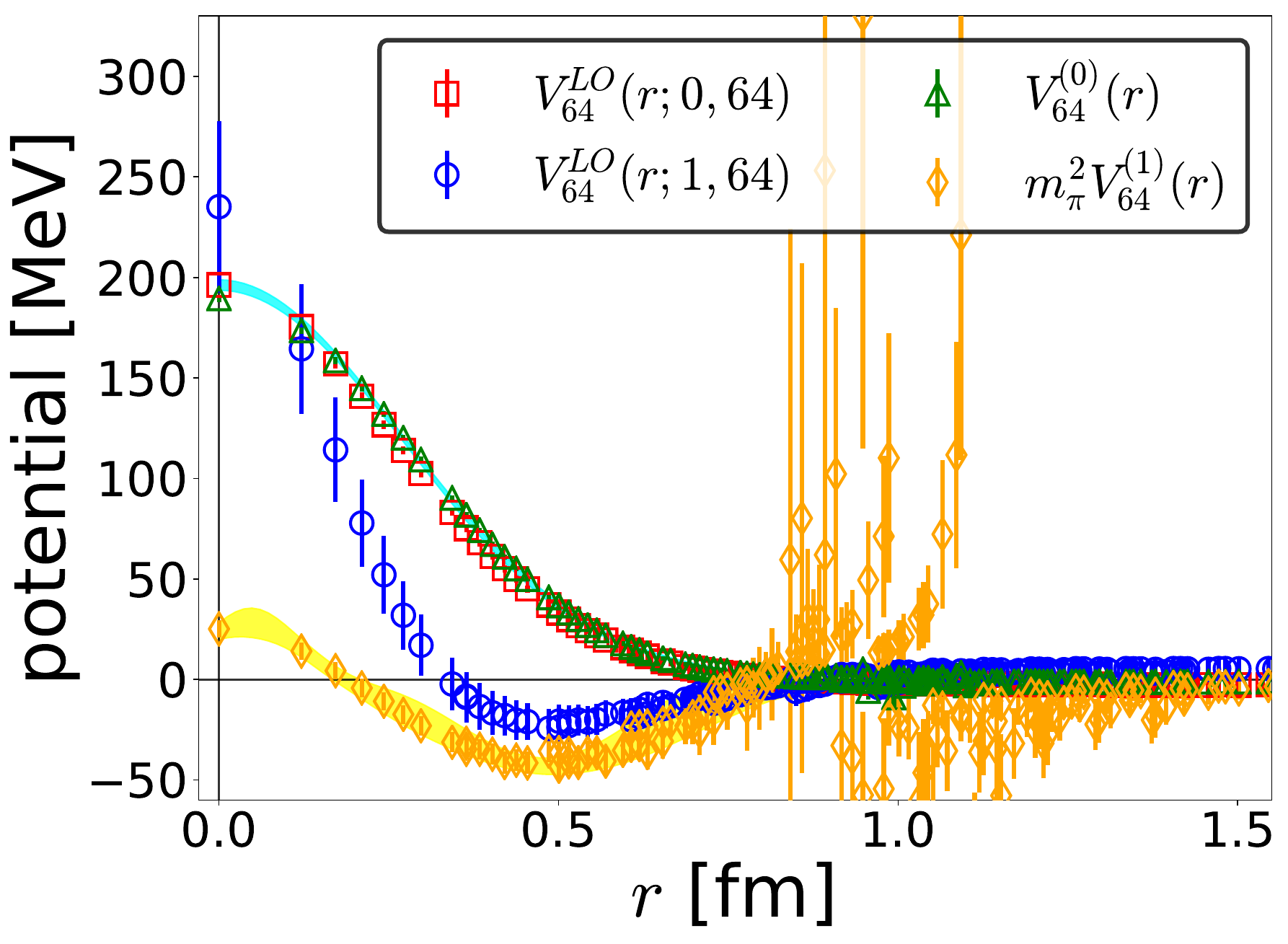}
\caption{The result of the NLO analysis, $V^{(0)}_{64}(r)$ (green up triangles) and $m_\pi^2V^{(1)}_{64}(r)$ (orange diamonds),
  together with $V^{\rm LO}_{64}(r; 0,64)$ (red squares) and $V^{\rm LO}_{64}(r; 1,64)$ (blue circles) for a comparison.
 Light blue and yellow bands correspond to fit results of $V^{(0)}_{64}(r)$ and $V^{(1)}_{64}(r)$, respectively. 
}
\label{fig:lonlo_ssink}
\end{figure}

We now consider the potential in the smeared-sink scheme.
Fig.~\ref{fig:tot_pot_ssink} shows the effective LO potentials in the smeared-sink scheme with $n_a=64$ from the smeared source with $n_b=64$, $\vert \bold{P} \vert = 0,1$, obtained at $t-t_0 = 11$. It turns out that potentials have sizable source momentum dependence unlike the case of the point-sink scheme.
It is also found that the potentials have non-negligible dependence on $t-t_0$ for $\vert \bold{P} \vert = 1$.
These observations indicate that the NLO contribution cannot be neglected in this scheme. 
We therefore determine $V^{(0)}$ and $V^{(1)}$ separately, by solving  
\begin{equation}
\label{eq:singular}
\begin{pmatrix}
1 & \dfrac{\nabla^2 R^{A^+_1,1}_{n_a,n_b}(\bold{r}, t; 0,t_0)}{R^{A^+_1,1}_{n_a,n_b}(\bold{r}, t; 0,t_0)} \\
\\
1 & \dfrac{\nabla^2 R^{A^+_1,1}_{n_a,n_b}(\bold{r}, t; 1,t_0)}{R^{A^+_1,1}_{n_a,n_b}(\bold{r}, t; 1,t_0)}
\end{pmatrix}
\begin{pmatrix}
V^{(0)}_{n_a}(r) \\
\\
V^{(1)}_{n_a}(r)
\end{pmatrix}
= 
\begin{pmatrix}
V^{\rm LO}_{n_a}(r; 0,n_b) \\
\\
V^{\rm LO}_{n_a}(r; 1,n_b).
\end{pmatrix}.
\end{equation}
Fig.~\ref{fig:lonlo_ssink} shows $V^{(0)}_{64}(r)$ (green up triangles) and $m_\pi^2V^{(1)}_{64}(r)$ (orange diamonds), together with
$V^{\rm LO}_{64}(r;0,64)$ (red squares) and $V^{\rm LO}_{64}(r;1,64)$ (blue circles).
One first notices $V^{(0)}_{64}(r) \simeq V^{\rm LO}_{64}(r;0,64)$, which suggests 
$V^{(1)}_{64}(r) \nabla^2 R^{A^+_1,0}_{64,64} (\bold{r}, t; 0, t_0) $ is negligibly small.
This means that 
nonzero momentum components in $R^{A^+_1,0}_{64,64} (\bold{r}, t; 0, t_0)$ is tiny.
On the other hand, $R^{A^+_1,0}_{64,64} (\bold{r}, t; 1, t_0)$ contains
non-zero momentum component enough to determine  $V^{(0)}_{64}(r)$ and $V^{(1)}_{64}(r)$ separately, the latter of which is responsible for the difference between $V^{\rm LO}_{64}(r;0,64)$ and $V^{\rm LO}_{64}(r;1,64)$ in Fig.~\ref{fig:tot_pot_ssink}.

\subsection{Scattering phase shifts in the smeared-sink scheme}
\label{subsec4}
We obtain the scattering phase shifts by solving the Schr\"{o}dinger equation with the potential
including the NLO term.
For the NLO analysis, $V_{n_a}^{(0)}(r)$ is fitted  with the  form 
\begin{equation}
V^{(0)}(r) = A_1 e^{-\frac{r^2}{\sigma_1^2}} + A_2 e^{-\frac{r^2}{\sigma_2^2}}
\label{eq:pot_LO}
\end{equation}
for all $r$, while
$V_{n_a}^{(1)}(r)$ is fitted with
\begin{equation}
V^{(1)}(r) = A_3 e^{-\frac{r^2}{\sigma_3^2}} + A_4 e^{-\frac{(r-a)^2}{\sigma_4^2}}
\label{eq:pot_NLO}
\end{equation}
for  $0 \le r  \le 0.8$ fm, which is chosen so as to exclude singular behaviors of $V_{n_a}^{(1)} (r)$ at large $r$ in the fit.
In Fig.~\ref{fig:lonlo_ssink}, fitted functions for $V^{(0)}(r)$ and  $V^{(1)}(r)$ are also given by light-blue and yellow bands, respectively.

\begin{figure}[htbp]
\includegraphics[clip, width=\textwidth]{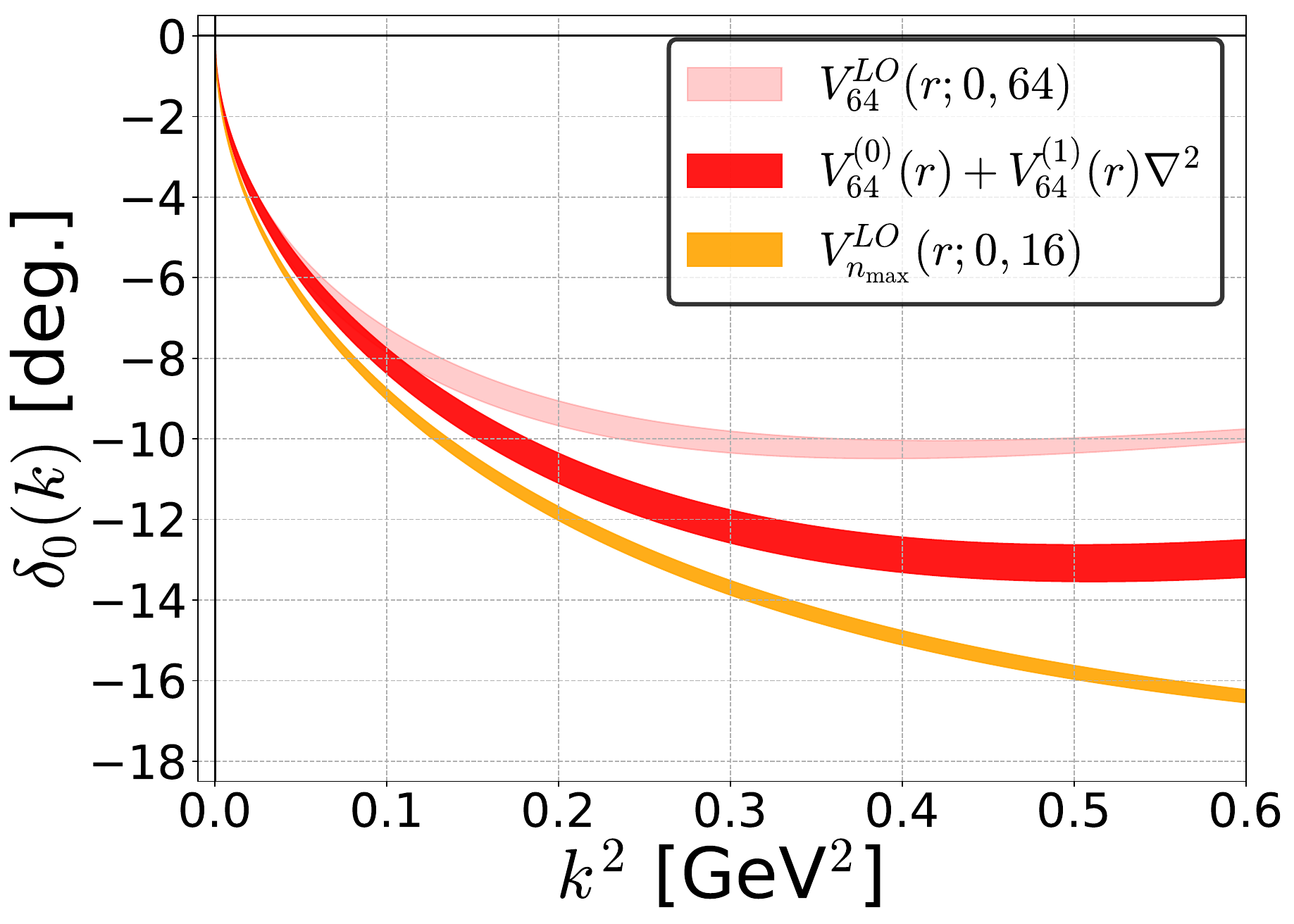}
\caption{The phase shifts of the S-wave $I=2$ $\pi\pi$ scattering from the potential 
in the point-sink scheme (LO: orange) and the smeared-sink scheme (LO: pink, NLO: red)
as a function of $k^2$.
}
\label{fig:phase_lonlo}
\end{figure}
In order to see an effect of the NLO correction on  the scattering phase shifts $\delta_0(k)$ quantitatively,
we show the scattering phase shifts as a function of  $k^2$ in Fig.~\ref{fig:phase_lonlo},
where the benchmark result from $V^{\rm LO}_{n_{\rm max}}(r;0,16)$ (orange) in the point-sink scheme is compared with those from $V^{\rm LO}_{64}(r;0,64)$ (pink) and  $V^{(0)}_{64}(r) + V^{(1)}_{64}(r)\nabla^2$ (red) in the smeared-sink scheme.

Let us first compare the scattering phase shifts between the point-sink scheme (orange) and the smeared-sink scheme (pink) at the LO.
While both results agree with each other at very low energies, they start deviating as the energy increases. The difference in the phase shift becomes about 4 degrees at $k^2 = 0.3$ GeV$^2$. 
If the NLO analysis is employed in the smeared-sink scheme, the agreement between the point-sink scheme (orange) and the smeared-sink scheme (red) is improved as expected.
The difference in the phase shift is reduced to about 2 degrees ($\simeq 15$ \%) at $k^2 =0.3$ GeV$^2$.
This result indicates that the use of the NLO analysis is almost mandatory in the smeared-sink scheme for quantitatively precise descriptions of scattering phenomena by the HAL QCD potential method.
 
\begin{figure}[htbp]
\includegraphics[clip, width=\textwidth]{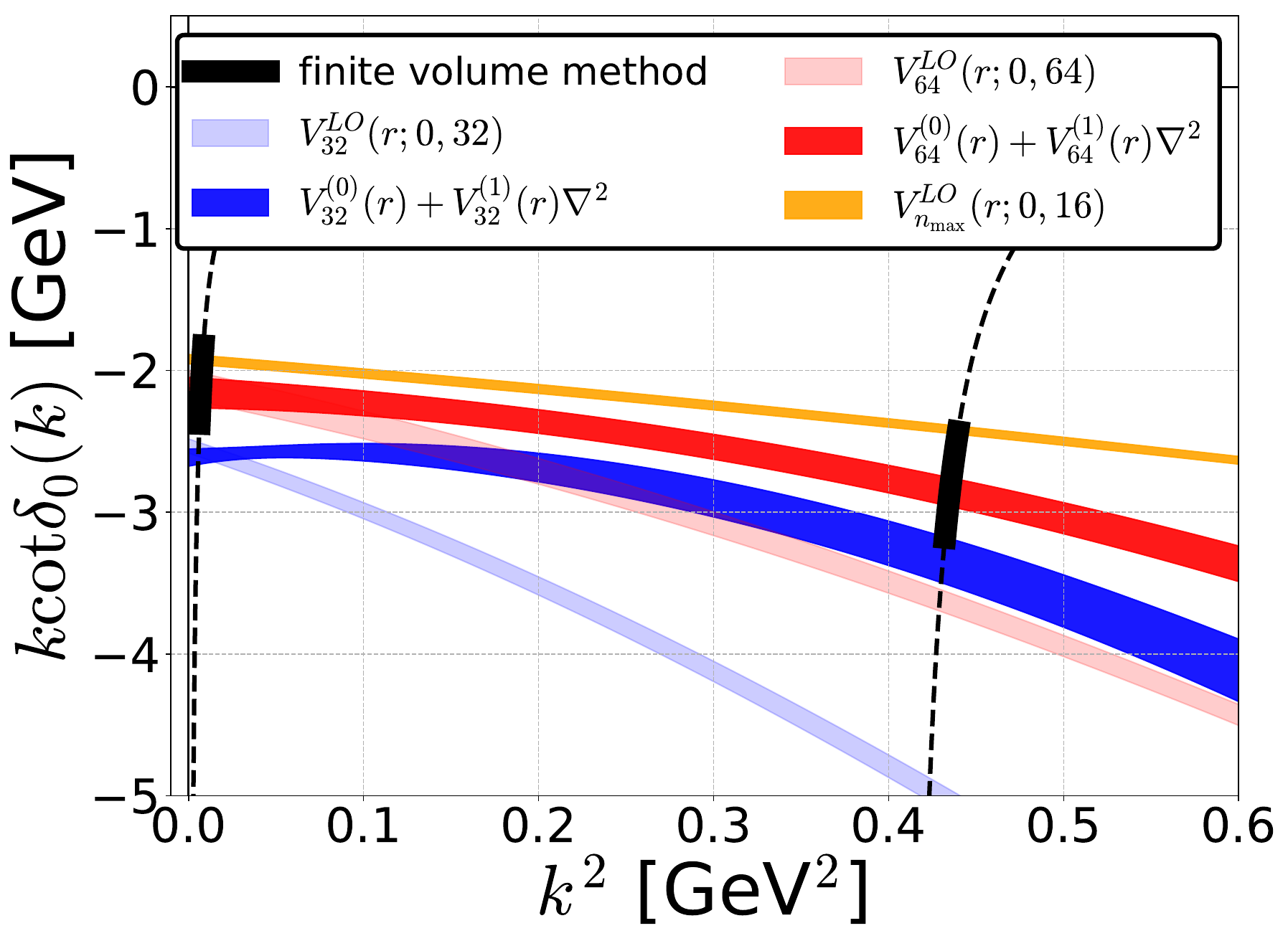}
\caption{$k \cot \delta_0(k)$ as a function of $k^2$ for the point-sink scheme (orange), the smeared-sink scheme with $n_a=n_b=64$ (LO: pink, NLO: red) and with $n_a=n_b=32$ (LO: light blue, NLO: blue).
Black bands show the phase shift calculated from the finite volume energy shift through the 
L\"{u}scher's formula, $2 \mathcal{Z}_{00}(1;(\frac{kL}{2\pi})^2)/\pi^{\frac{1}{2}} L$, denoted by black dashed lines.
}
\label{fig:kcot_lonlo}
\end{figure}

In order to see differences more precisely, $k \cot \delta_0(k)$ is plotted as a function of $k^2$
in Fig.~\ref{fig:kcot_lonlo}, 
where
the phase shifts obtained from the finite volume energy by the L\"{u}scher's finite volume method \cite{luscher} (see Appendix.~\ref{sec:var} for details) are also shown
at $k^2 = 7.2 \times 10^{-3}(_{-5}^{+13})$ and $0.43(^{+1}_{-0})$ GeV$^2$ (black bands). 
It is clearly observed that
the NLO analysis significantly improves the agreement between different schemes/methods.
The results from the effective LO potential in the point-sink scheme (orange)
as well as those from the NLO potentials in the smeared-sink schemes with $n_a=64$ (red)
and $n_a=32$ (blue)
agree with results from the finite volume energy shift
not only at low energy ($k^2\simeq 0$ GeV$^2$) but also at higher energy ($k^2 \simeq 0.43$ GeV$^2$)
within their statistical errors, 
though statistical errors of $k\cot \delta_0(k)$ from the finite volume energy shift (black bands) are rather large.
Among different schemes in the potential method,
we remind readers that the results from the effective LO potential in the point-sink scheme is the most reliable as the benchmark,
since  the phase shifts in the point-sink scheme are robust in the sense that the source momentum dependence (the NLO term) is sufficiently small.
Therefore, the difference of results between the point and smeared-sink schemes gives the remaining systematic errors
in the smeared-sink scheme. The most plausible origin for this systematics at higher energies is the higher order corrections
in the derivative expansion in the smeared-sink scheme. In fact, the magnitude of the remaining systematics is
compatible with the magnitude of NNLO corrections which can be roughly estimated by the difference between LO and NLO results.
Another possible origin is the systematics in NLO terms, since there exist uncertainties in $V^{(1)}(r)$ at long range
($r \gtrsim 0.8$ fm in Fig.~\ref{fig:lonlo_ssink}).
Note that NLO (and higher order) corrections can be reduced by increasing the smearing level $n_a$,
as can be seen from the comparison between $n_a=32$ and 64.

At very low energies, $k^2\simeq 0$ GeV$^2$,
while the results between different schemes/methods achieve good agreement even at the LO analysis,
the inclusion of the NLO contribution does not resolve the remaining (small) deviations
between the point and smeared-sink schemes.
Considering that NLO correction is small and
that the derivative expansion is expected to be good at low energies,
higher order corrections would not be the origin of this systematics.
One of the possible reasons is the uncertainties in $V^{(1)}(r)$ at long range as mentioned above.
Another possibility is the systematics associated with the long tail structure in the LapH smearing (Fig.~\ref{fig:dist}),
and studies to improve the locality of the LapH smearing are in progress. 
Again, we can reduce the systematics by increasing the smearing level $n_a$, as is actually seen in the figure.

In Table \ref{tab:sigma}, we present $k\cot\delta_0(k)$ at $k^2 = 7.2\times 10^{-3}$ and 0.43 GeV$^2$, obtained from the effective LO potential
in the point-sink scheme ($n_a=n_{\rm max}$) with $\vert {\bf P}\vert =0$, as the benchmark result.
We also show the LO/NLO analyses in the smeared-sink scheme with $n_a=64$,
together with  results from the finite volume method,
where errors are statistical only.
Systematic errors in the smeared-sink scheme can be estimated by the
difference between the point and smeared-sink schemes as discussed above.
We then confirm that the potential method with the smeared-sink scheme
and finite volume method give consistent results with similar sizes of uncertainties,
which are dominated by systematic errors in the former but by statistical errors for the latter.

\begin{table}[tbhp]
\begin{tabular}{|c||c|c|c||c|}
\hline
$k^2$ (GeV$^2$) & $V^{\rm LO}_{n_{\rm max}}(r;0,16)$ & $V^{\rm LO}_{64}(r;0,64)$  & $V^{(0)}_{64}(r) + V^{(1)}_{64}(r)\nabla^2$ &finite volume method \\
\hline
$7.2\times 10^{-3}$ & $-1.94(10)$  & $-2.13(10)$ & $-2.17(11)$ &$-2.11 (28)$  \\ \hline
$0.43$ & $-2.42(3)$ & $-3.65(8)$ &  $-2.86(11)$ &$-2.83(41)$ \\ \hline
\end{tabular}
\caption{
A comparison of $k\cot\delta_0(k)$ in units of GeV among various schemes/methods. Errors are statistical only.
}
\label{tab:sigma}
\end{table}

\section{Conclusion}
\label{sec:sec5}
In this paper,
the scheme independence in the HAL QCD method has been investigated for the $I=2$ $\pi\pi$ scattering phase shifts at $m_{\pi} \simeq 870$ MeV.
We have considered two schemes, the point-sink  scheme and the smeared-sink scheme with various smearing levels $n_a$, where we newly introduce the LapH smearing in the latter.
We have found in the point-sink scheme, which is the standard scheme in the HAL QCD method, 
that the NLO contributions are sufficiently small below $k^2 \sim 0.4$ GeV$^2$.
This means that the (effective) LO potential in the point-sink scheme is a good description of the $I=2$ $\pi\pi$ interaction at least up to $k^2 \sim 0.4$ GeV$^2$ ( $2.15$ GeV in terms of the central mass energy).\footnote{Note that the threshold of 4 $\pi$ in the final state corresponds to $k =\sqrt{3} m_\pi \simeq 1.5$ GeV.}
This result in full QCD confirms the conclusion in previous quenched QCD studies that the point-sink scheme is a good scheme for the HAL QCD potential for the $NN$ system \cite{spec6} and the $I=2$ $\pi\pi$ system \cite{I2pre}.  

In the case of the smeared-sink scheme,
the effective LO potential  shows a sizable dependence on the momentum at the source, so that the NLO contribution is non-negligible.
In fact, at the effective LO analysis, the scattering phase shifts in the smeared-sink scheme show some
deviation from the benchmark result given by the point-sink scheme, 
which increases as the energy increases.\footnote{In Appendix~\ref{sec:lo_analysis}, we present the effective LO analysis in the smeared-sink scheme with various smearing levels $n_a$.}
We thus calculated the phase shifts using the NLO analysis in the smeared sink 
scheme.
The NLO analysis improves an agreement with the benchmark result at both low and high energies,
and 
results in both point-sink scheme and smeared-sink scheme with $n_a=64$
are consistent with the scattering phase shifts from the finite volume energy shift calculated by the L\"uscher's  formula at two energies,
within large statistical errors in the finite volume method.
The systematic errors of the potential method in the smearing sink scheme with $n_a = 64$
  can be estimated from the difference between the point-sink scheme and the smeared-sink scheme,
  and are found to be similar size compared to the statistical errors of the finite volume method.

Our study in this paper gives the following lessons for future studies of resonant states such as $I=0, 1$ $\pi\pi$ scatterings
by the HAL QCD potential method in the smeared-sink scheme.
(1) It is better to increase the smearing level $n_a$ as long as computational resources (cpu time, memory, disk space, etc.) allow.
(2) It is better to observe the $n_a$ dependence carefully.
(3) It is important to employ the NLO  analysis if possible.

Studies of resonant states in the HAL QCD potential method 
are ongoing. Results will be reported in near future.

\section{acknowledgement}
This work is supported in part by
the Grant-in-Aid of the Japanese Ministry of Education, Sciences and Technology, Sports and Culture (MEXT) for Scientific Research (No.~JP16H03978, (C)26400281, JP15K17667),
by a priority issue (Elucidation of the fundamental laws and evolution of the universe) to be tackled by using Post ``K" Computer, 
and by Joint Institute for Computational Fundamental Science (JICFuS).
D.K. is supported in part by the Japan Society for the Promotion of Science (JSPS).
T.D. is supported in part by RIKEN iTHES Project and iTHEMS Program.
We thank the JLQCD collaboration and CP-PACS collaboration for providing us their 2+1 flavor gauge configurations \cite{conf1, conf2}.
The code for the eigenvectors of gauge covariant Laplacian was supplied by Colin Morningstar. We also thank for his kindness.
All the numerical calculations are done on the Cray XC40 in Yukawa Institute for Theoretical Physics (YITP) in Kyoto University.

\appendix
\section{The maximum eigenvalue and the spatial distribution  vs. the smearing level}
\label{sec:k2}
\begin{figure}[htbp]
\includegraphics[clip,width=\textwidth]{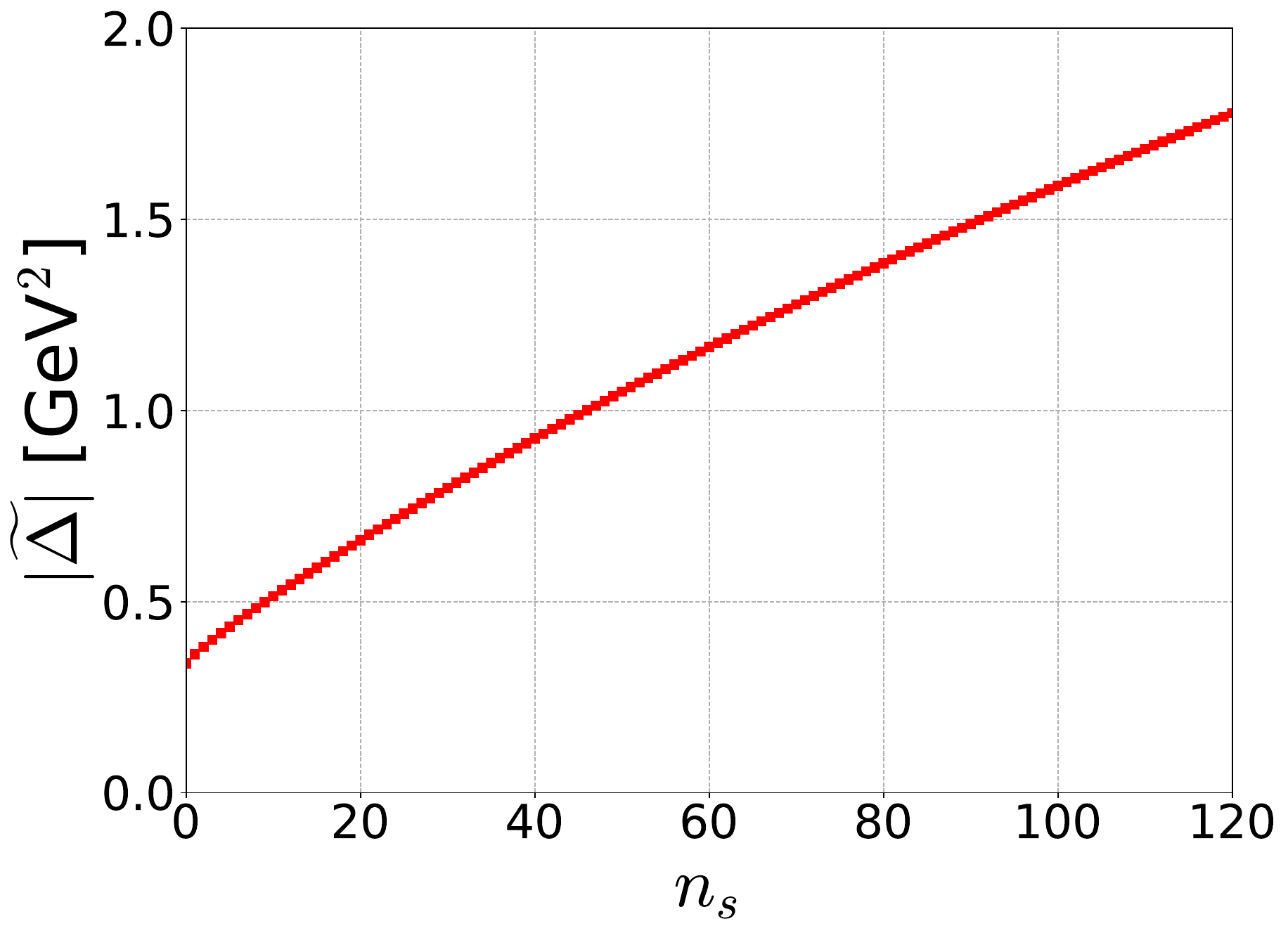}
\caption{The maximum eigenvalue of lattice covariant Laplacian $\vert \tilde{\Delta} \vert$ as a function of $n_s$.  }
\label{fig:eigval}
\end{figure}
Fig.~\ref{fig:eigval} shows the maximum eigenvalue of the lattice covariant Laplacian in Eq.~(\ref{eq:laplacian}), $\vert \tilde{\Delta} \vert$ [GeV$^2$],  as a function of the  smearing level $n_s$.

\begin{figure}[htbp]
\includegraphics[clip,width=\textwidth]{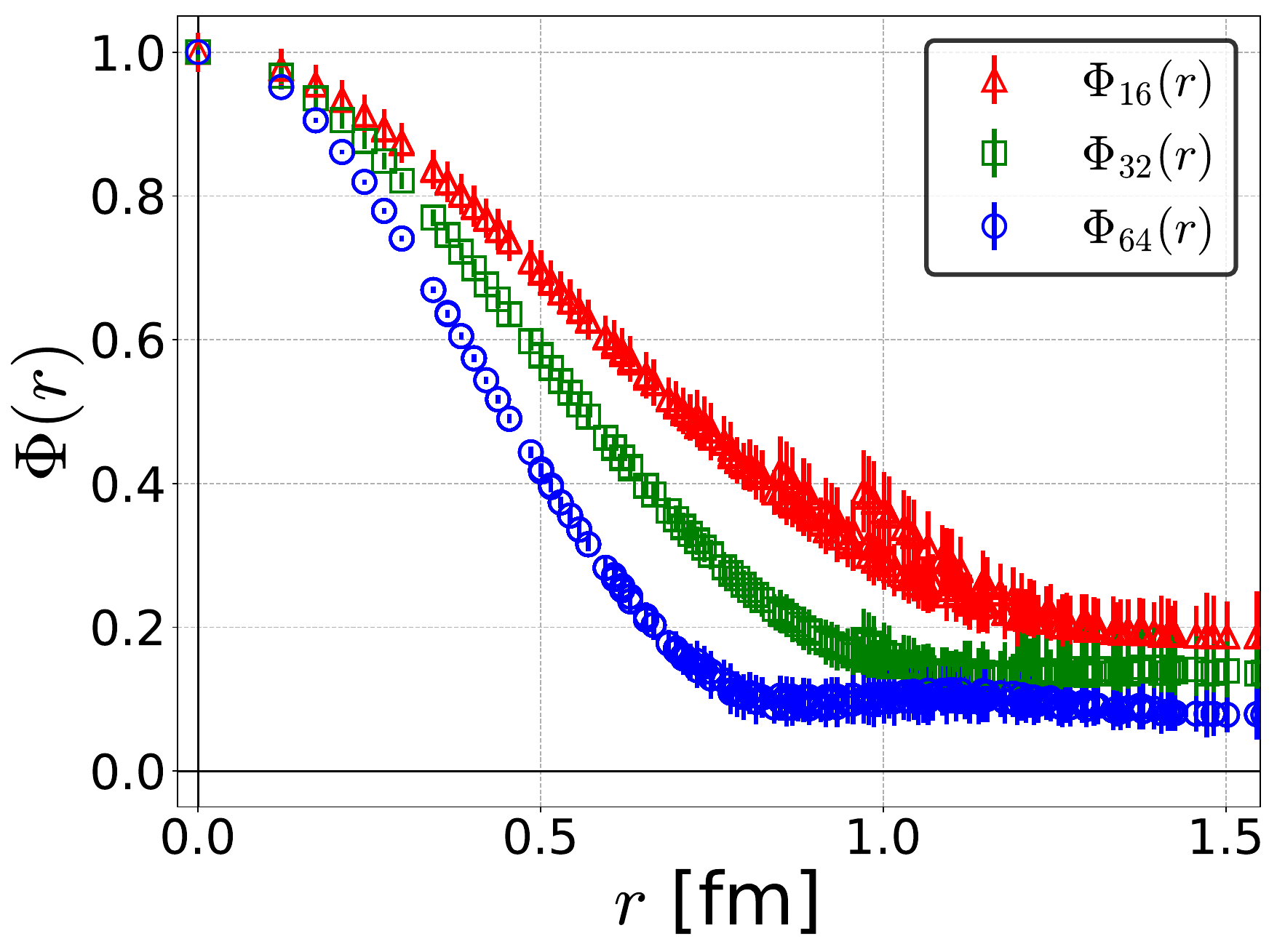}
\caption{A measure for the spatial distribution of LapH smearing operator.}
\label{fig:dist}
\end{figure}

In Ref.~\cite{distill}, a measure of  the spatial distribution for the LapH smearing is defined by
\begin{equation}
  \Phi_{n_s}(\bold{r}) = \sum_{\bold{x},t} \sqrt{ \mathrm{Tr} \left\{ S_{n_s}(\bold{x},\bold{x+r},t) S_{n_s}(\bold{x+r},\bold{x},t) \right\} } ,
\end{equation}
which is gauge invariant and represents to what extent quark field is smeared.
Fig.~\ref{fig:dist} gives this quantity  for $n_s = 16, 32,$ and $64$.
The figure tells us that the smeared quark is more localized as $n_s$ increases.

\section{Finite volume energies from the variational method}
\label{sec:var}
In the calculation of the finite volume energy,
unwanted constant contribution from thermal quark loops
to the 2-pion correlation function are removed as in Ref.~\cite{hadI2},
\begin{eqnarray}
\widetilde{\mathcal{C}}^{4,\Lambda, \mu}_{n_a,n_b} (\bold{r},t;\left|\bold{P}\right|,t_0)  &=& \mathcal{C}^{4,\Lambda, \mu}_{n_a,n_b} (\bold{r},t;\left|\bold{P}\right|,t_0) \nonumber \\
 &-& \mathcal{C}^{4,\Lambda, \mu}_{n_a,n_b} (\bold{r},t  + \Delta t;\left|\bold{P}\right|,t_0),
\end{eqnarray}
where we set $\Delta t = 1$ in this paper.

\begin{figure}[htbp]
\includegraphics[clip,width=0.85\textwidth]{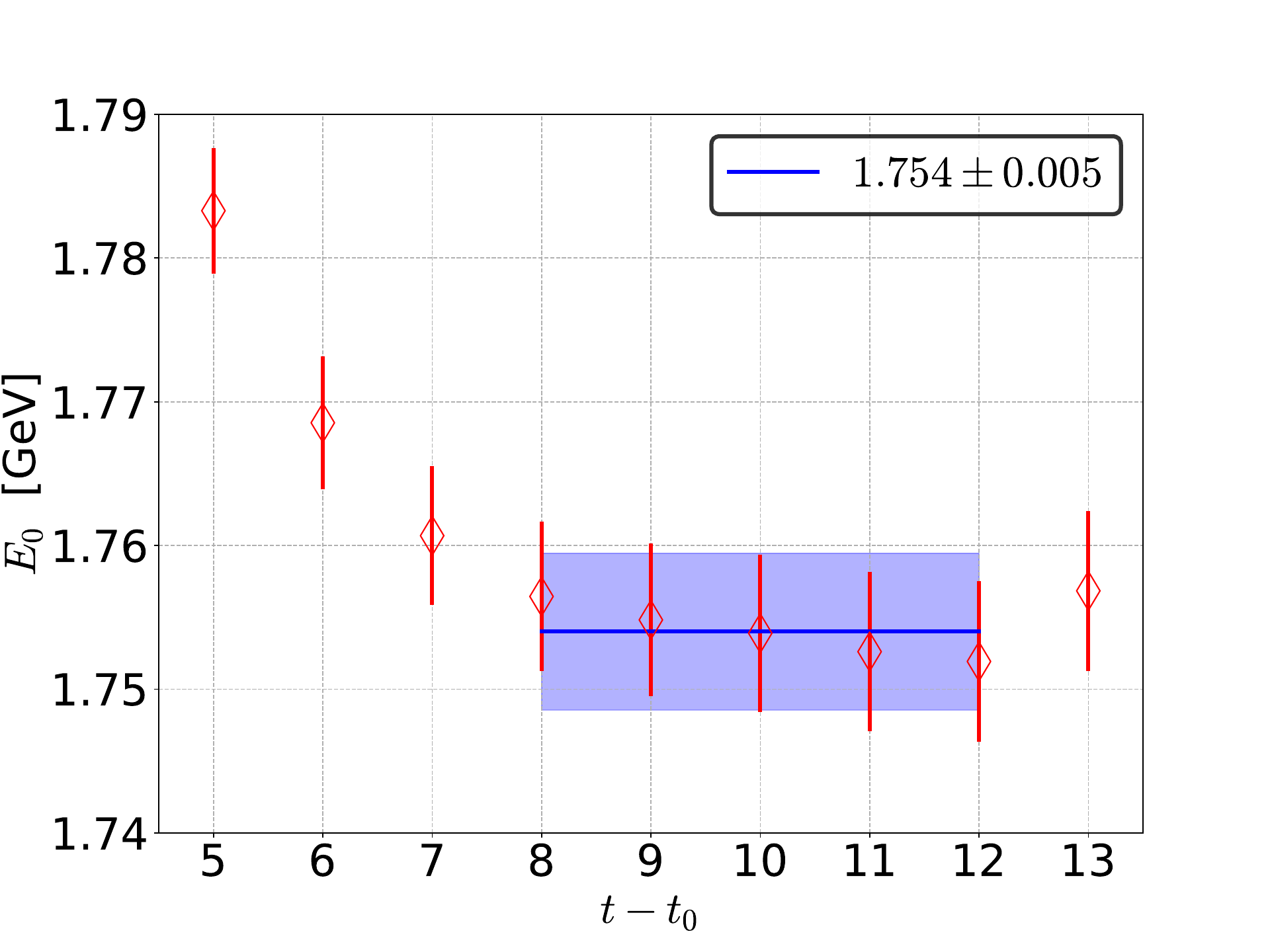}
\includegraphics[clip,width=0.85\textwidth]{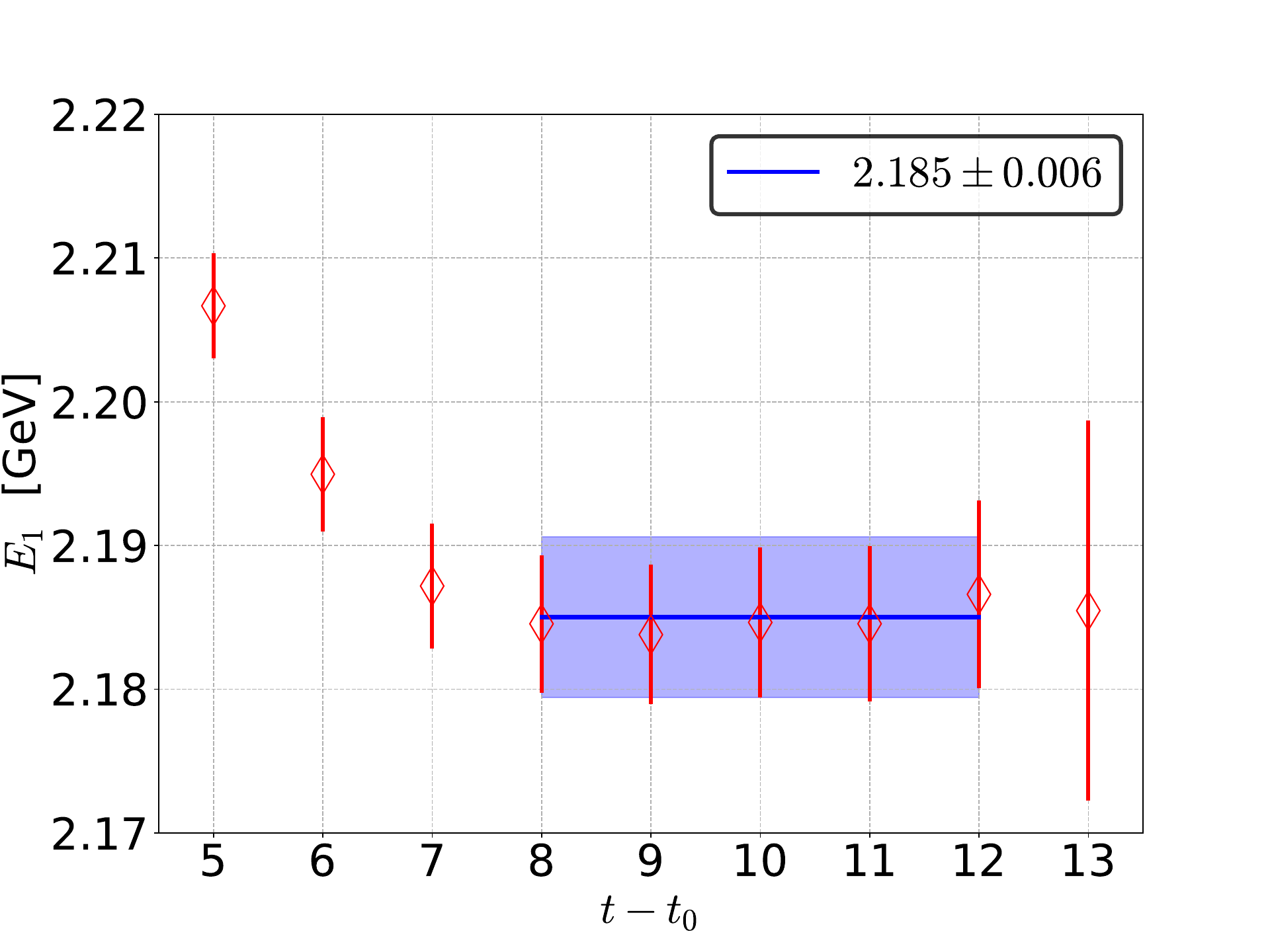}
\caption{The effective energy given by  the variational method for $\widetilde{\mathcal{C}}^{4,A^+_1, 1}_{n_a,n_b} (\bold{r},t;\left|\bold{P}\right|,t_0)$.
Upper and lower panel show the ground state energy and the first excited state energy, respectively.
Blue solid lines with blue bands represent central values and errors from fits with data in these intervals.
 }
\label{fig:eff_energy}
\end{figure}
In the variational method \cite{vari1,vari2}, we consider the matrix
\begin{eqnarray}
\label{eq:corr_mat}
\mathcal{C}(P,t,P',t_0) & = &  C(P,P_1;t_0)^{-\frac{1}{2}}  \\
&& C(P_1,P_2;t) C(P_2,P'; t_0)^{-\frac{1}{2}}, \nonumber
\end{eqnarray}
where $C(P,P';t)$ is given by the Fourier transformed 4-pt correlator
\begin{equation}
C(P,P';t) = \sum_{t_1,t_2 \atop t = t_1-t_2} \sum_{P \atop |P| : {\rm fixed}} \sum_{\bold{r}} e^{-i\bold{P}\cdot \bold{r}} \widetilde{\mathcal{C}}^{4,A^+_1,0}_{n_a,n_b}(\bold{r},t_1;|\bold{P'}|,t_2).
\end{equation}
We denote $\lambda_n(t,t_0)$ as the eigenvalue of the matrix in Eq.~(\ref{eq:corr_mat}).
The eigenenergy of the system $E_n$ is extracted as
\begin{equation}
E_n = \lim_{t \to \infty} E_n(t), \qquad
E_n(t-t_0) \equiv -\frac{1}{t-t_0} \log \lambda_n(t,t_0) ,
\end{equation}
where we assume $E_0 \leq E_1 \leq E_2 \leq \dots $ and $E_n(t)$ is called the effective energy.
In this study, we employ a $3 \times 3$ matrix from $P,P' = 0, 1, 2$ correlation functions with $n_a=n_b= 64$. 
Fig.~\ref{fig:eff_energy} shows the effective energy $E_{0,1}(t-t_0)$ as a function of $t-t_0$.

Once $E_n$'s are obtained, 
phase shifts can be extracted  by the L\"{u}scher's finite volume method \cite{luscher}.
We calculate the momentum $k^2$, which corresponds to the energy difference from the 2 pion mass,
$k^2_n = \frac{E_n^2}{4} - m_\pi^2$.
As long as $ l \ge 4$ partial waves are negligible,
the L\"{u}scher's formula for $A_1^+$ representation relates
$k^2_n$  to the scattering phase shift as 
\begin{equation}
k_n \cot \delta_0(k_n) = \frac{2 \mathcal{Z}_{00}(1;q_n^2)}{\pi^{\frac{1}{2}} L},
\end{equation}
where $L$ is the spatial lattice extension, $q_n$ is the dimensionless momentum defined as $q_n~\equiv~k_nL/2\pi$ and $\mathcal{Z}_{00}$ is a generalized zeta function, $\mathcal{Z}_{00}(s,q^2) \equiv \frac{1}{\sqrt{4\pi}} \sum_{\bold{n} \in \mathbb{Z}^3} \left(\bold{n}^2 - q^2 \right)^{-s}$.

\section{Effective LO analysis in the smeared-sink scheme} 
\label{sec:lo_analysis}
  As discussed in the main text (Sec.~\ref{subsec3}),
  NLO correction is large in the smeared-sink scheme in particular at high energies,
  and the effective LO analysis is not sufficient.
  In this appendix, we nonetheless present the effective LO analysis in the smeared-sink scheme
  and demonstrate how the truncation error in the derivative expansion appears in the lattice QCD results.

\subsection{Effective LO potentials}
\label{subsec:smeared}
\begin{figure}[htbp]
\includegraphics[clip, width=\textwidth]{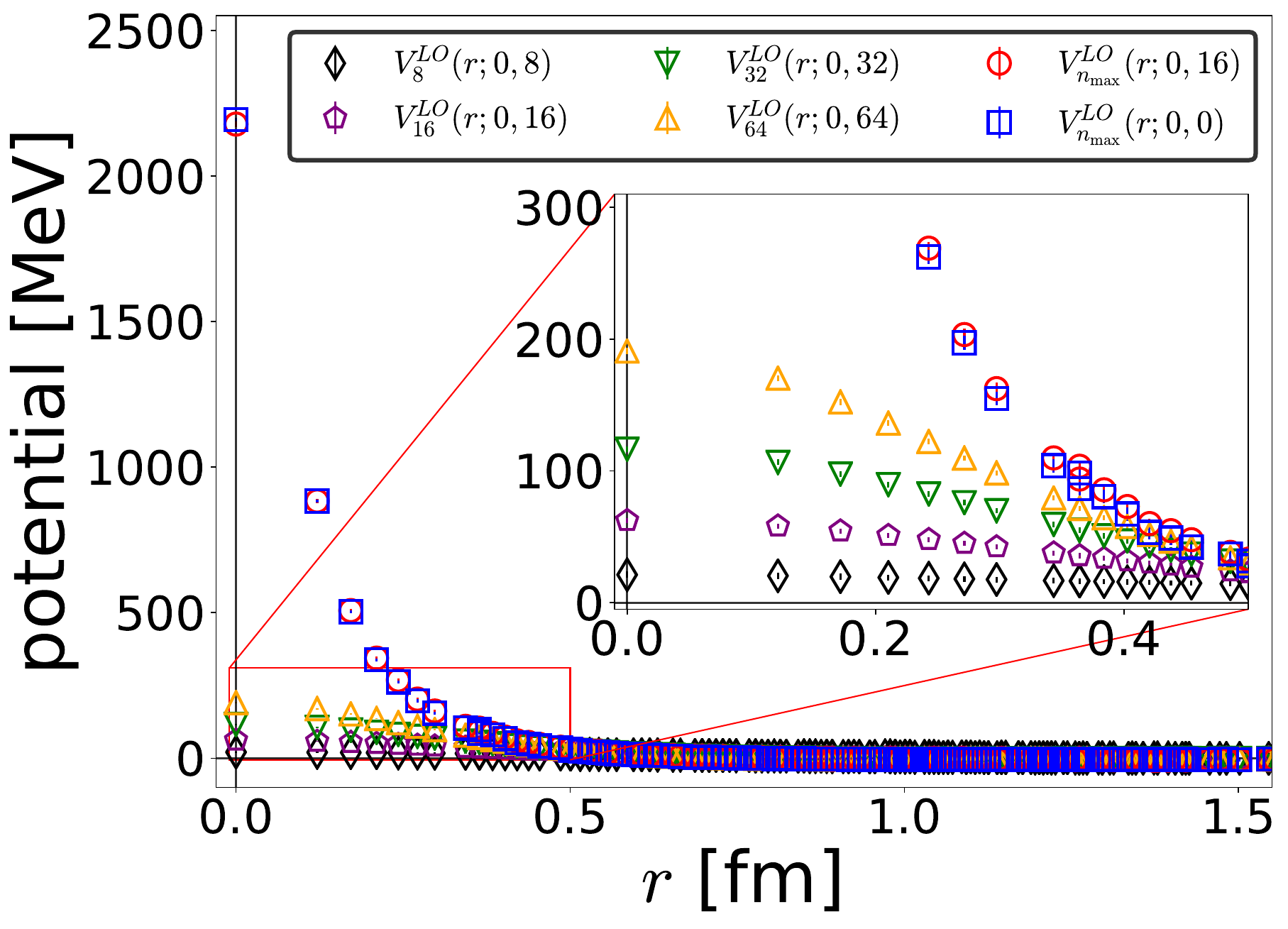}
\caption{
 The effective LO potential $V^{\rm LO}_{n_a}(r;0,n_b)$ for several combinations of $(n_a,n_b)$.
    Black diamonds, purple pentagons, green down triangles and orange up triangles correspond to $V^{\rm LO}_{8}(r;0,8)$, $V^{\rm LO}_{16}(r;0,16)$, $V^{\rm LO}_{32}(r;0,32)$ and $V^{\rm LO}_{64}(r;0,64)$, respectively, while red circles and blue squares give $V^{\rm LO}_{n_{max}}(r;0,16)$ and $V^{\rm LO}_{n_{max}}(r;0,0)$, respectively.
The inset shows the comparison of the repulsion at short distance.
}
\label{fig:leading_pot}
\end{figure}
  Fig.~\ref{fig:leading_pot} shows the effective LO potential in the smeared-sink scheme $V_{n_a}^{\rm LO}(r; 0, n_b)$ at $\bold{P} = [0,0,0]$,
 together with   $V_{\rm n_{max}}^{\rm LO}(r; 0, n_b)$ for a comparison. 
  As seen from the figure, effective LO potentials in the smeared-sink scheme have much reduced repulsive core than those in the point-sink scheme, while the long-distant part of the potential looks similar for all cases including the one in the point-sink scheme.
We remind the readers that the potential is expected to be dependent on the scheme.
  In the inset of Fig.~\ref{fig:leading_pot}, a short distance part of potentials in the  smeared-sink scheme is also shown. The magnitude of the potential at short distance increases monotonically as $n_a$  increases (equivalently the size of the smearing range decreases as shown in Appendix.~\ref{sec:k2}).

\subsection{Scattering phase shift}
\label{subsec:phase_s_lo}
\begin{figure}[tbhp]
    \includegraphics[clip,width=\textwidth]{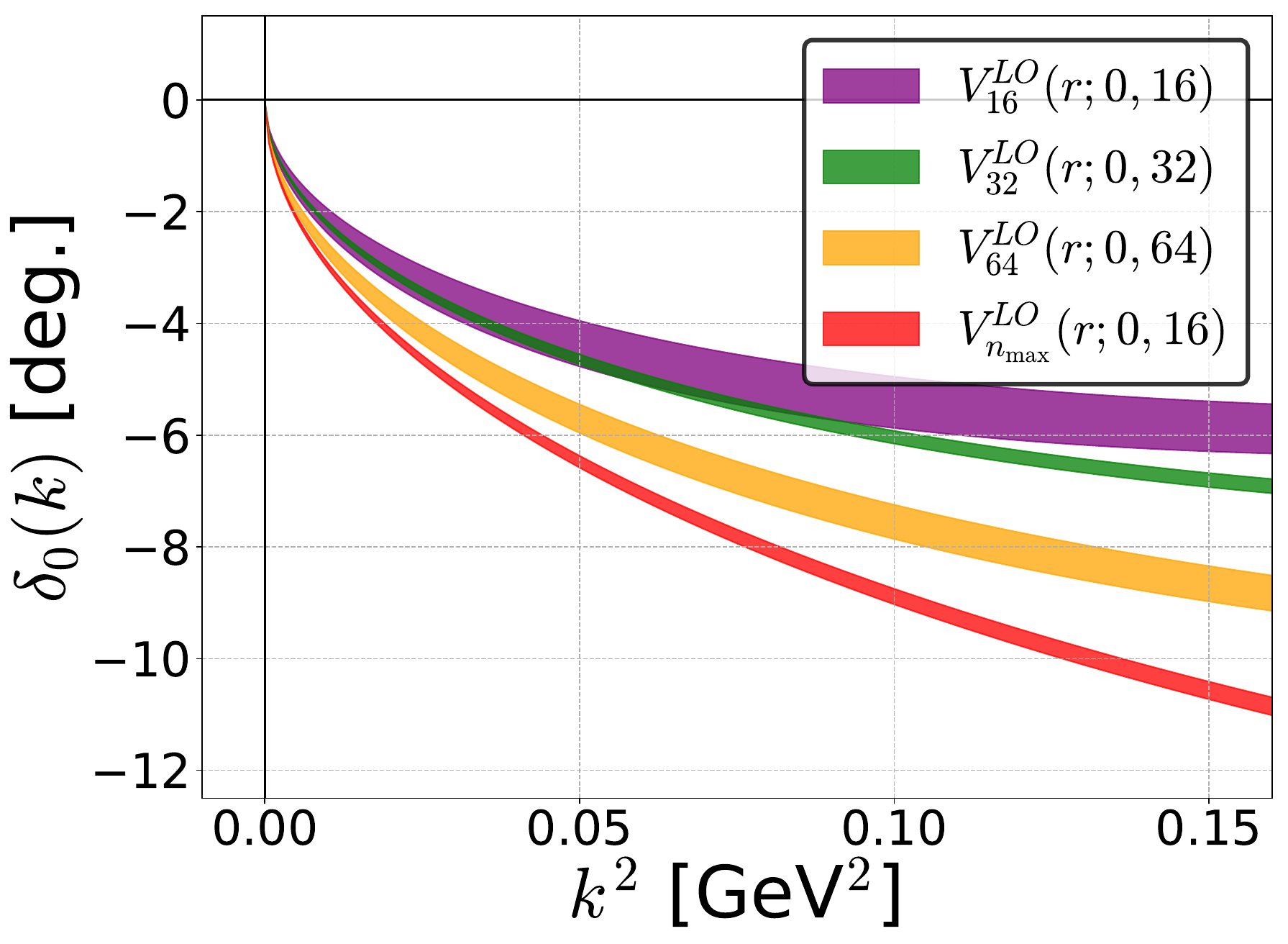}
    \caption{
      The phase shifts of the S-wave $I=2$ $\pi\pi$ scattering from the potential 
      in the point-sink scheme (red) and the smeared-sink scheme
      with $n_a=n_b=16$ (purple), 32 (green) and 64 (orange)
      as a function of $k^2$.
    }
    \label{fig:phase_static}
  \end{figure}

We study the S-wave $I=2$ $\pi\pi$ scattering phase shifts $\delta_0(k)$
in the effective LO analysis.
  We extract the scattering phase shifts by solving the Schr\"{o}dinger equation as described in Sec.~\ref{sec:scheme}.
   We fit potentials in Fig.~\ref{fig:leading_pot} with
  \begin{equation}
    V^{\rm LO}(r) = A_1 e^{-\frac{r^2}{\sigma_1^2}} + A_2 e^{-\frac{r^2}{\sigma_2^2}}
  \end{equation}
  for all $r$.
  Fig.~\ref{fig:phase_static} shows the phase shifts obtained from $V^{\rm LO}_{16}(r;0,16)$ (purple), $V^{\rm LO}_{32}(r;0,32)$ (green) and $V^{\rm LO}_{64}(r;0,64)$ (orange), together with
the benchmark result from $V^{\rm LO}_{n_{\rm max}}(r;0,16)$ (red), as a function of $k^2$, where $k$ is the magnitude of the scattering momentum.
 
Phase shifts in Fig.~\ref{fig:phase_static} show non-negligible dependence on the sink operator scheme.
As the smearing level $n_a$ increases, the phase shifts show more repulsive behavior and approach to the benchmark result in the point-sink scheme. 
The difference of the phase shifts between the smeared-sink scheme and the point-sink scheme is small at low energies but gradually become larger as $k^2$ increases.
The discrepancy in phase shifts between the smeared sink and the point sink is originated from
sizable NLO contributions in the smeared-sink scheme, which however decreases as $n_a$ increases.


\end{document}